\def\be{\begin{equation}}
\def\te{\end{equation}}
\def\bea{\begin{eqnarray}}
\def\nn{\nonumber}
\def\tea{\end{eqnarray}}

\def\ha{{1\over 2}}

\def\intf{\int_0^\infty}
\def\intx{\int d^Dx \sqrt{-g}}
\def\intxf{\int d^4x \sqrt{-g}}
\def\intxt{\int d^2x \sqrt{-g}}
\def\met{g_{\mu\nu}}
\def\metu{g^{\mu \nu}}

\def\ricu{R^{\mu \nu }}

\def\einu{G^{\mu \nu}}

\def\xib{\overline{\xi}}
\def\rimt{R^{\alpha \beta \gamma \delta}R_{\alpha \beta \gamma \delta}}

\def\rimu{R^{\alpha \beta \gamma \delta}}

\def\a{\alpha}
\def\b{\beta}
\def\d{\delta}
\def\e{\epsilon}

\def\g{\gamma}

\def\k{\kappa}
\def\l{\lambda}
\def\m{\mu}
\def\mt{\tilde{\mu}}
\def\n{\nu}

\def\q{\theta}

\def\s{\sigma}
\def\t{\tau}

\def\D{\Delta}

\def\G{\Gamma}

\def\L{\Lambda}
\def\O{\Omega}

\def\bb{\bibitem}

\documentstyle[aps]{revtex}
\input{epsf}
\begin{document}
\title{Non-perturbative effects of vacuum energy on the recent expansion of the universe}
\author{Leonard Parker\thanks{Electronic address:
leonard@uwm.edu} and Alpan Raval\thanks{Electronic address: raval@uwm.edu}\\
{\small Department of Physics, University of Wisconsin-Milwaukee, P.O. Box 
413,
 Milwaukee, WI 53201.}}
\maketitle
\begin{abstract}
\noindent We show that the vacuum energy of a free quantized field of
very low mass can significantly alter the recent expansion of the
universe. The type of particle we consider is of spin-$0$, but a
higher spin field, such as a graviton of ultralight mass, may well
affect the expansion in the same way.
The effective action of the theory is obtained from a non-perturbative sum
of scalar curvature terms in the propagator. We renormalize
 this effective action and express it in terms of observable
gravitational coupling constants. We numerically investigate the
semiclassical Einstein equations derived from it. As a result of
non-perturbative quantum effects, the scalar curvature of the
matter-dominated universe stops decreasing and approaches a constant
value. The universe in our model evolves from an open
matter-dominated epoch to a mildly inflating de Sitter expansion. The
Hubble constant during the present de Sitter epoch, as well as the
time at which the transition occurs from matter-dominated to de Sitter
expansion, are determined by the mass of the field and by the
present matter density. 
The model provides a theoretical explanation of the
observed recent acceleration of the universe, and gives a good fit to
data from high-redshift Type Ia supernovae, with a mass of about
$10^{-33}$ eV, and a current ratio of matter density to critical density,
$\O_0 <0.4$. The age of the universe
 then follows
with no further free parameters in the theory, and
turns out to be greater than $13$ Gyr. 
The model is spatially open and
consistent with the possibility of inflation in the very early
universe. Furthermore, our model arises from the standard
renormalizable theory of a free quantum field in curved spacetime, and
does not require a cosmological constant or the associated
fine-tuning.
\smallskip
\noindent WISC-MILW-99-TH-6\\
April 28, 1999.

\end{abstract}
\newpage
\section{Introduction}

There appear to be deep connections between phenomena at the
microscopic quantum level and those at the macroscopic level described
by general relativity. These connections were first elucidated by
studies of quantum field theory in the classical curved spacetime of
general relativity, particularly those involving particle creation by
strong gravitational fields, such as exist in the early universe and
near black holes \cite{parker,hawking}. The framework of quantum field
theory in the classical curved spacetime of general relativity appears
to be quite robust, and is the one we employ in our present study. We
are particularly concerned with certain non-perturbative gravitational
contributions to the vacuum energy of quantized fields \cite{partom}.

We show that these contributions may account for the recent 
observations \cite{perl2,perl} of type-Ia supernovae (SNe-Ia) that seem 
to imply that there is an acceleration of the recent expansion 
of the universe. In our model, the key new ingredient needed to account for the
observations is the existence of a particle having a very small mass 
of about $10^{-33}$ eV. This could be a scalar particle or one of
higher spin, such as a graviton. It is well-known that in
a 
Robertson-Walker 
universe, the linearized Einstein equations obeyed by the 
graviton field take the same form (for each polarization in the 
Lifshitz gauge) as the equation of a minimally coupled massless 
scalar field \cite{lifshitz}. Inclusion of a mass of about 
$10^{-33}$ eV in these equations would give effects similar to the
ones studied here, and would evidently not
conflict with other observations. 
We consider a scalar field for
simplicity.

The effects we investigate stem from the discovery \cite{partom,parja}
 that a covariant infinite series of terms in 
the propagators of quantum fields in curved spacetime can be summed 
in closed form, to all orders in the curvature. The summed infinite series
of terms are {\em all} those that involve at least one factor of the scalar
curvature, R, (together with any number of factors of the Riemann tensor
and its covariant derivatives) in the Schwinger-DeWitt proper-time series 
for the propagator. We will refer to this partially-summed form of the 
propagator as the R-summed form of the propagator. 
Although only the first several terms in the proper-time series have
been calculated (because of their complexity), the R-summed form
can be derived by means of mathematical induction \cite{parja}.
A further reason to expect the R-summed form of the propagator to contain
information of physical significance, is that the leading term 
in the R-summed form of the propagator follows directly from 
the Feynman path-integral expression for the propagator by means of 
a Gaussian integration about the dominant path \cite{bekpa}.
Our results flow from the effective action that is obtained from the
R-summed form of the propagator.

We leave for a later paper, the contribution of these non-perturbative
gravitational effects to early inflation at the grand-unified scale
and associated particle creation. In this paper, we focus on 
possible consequences of these non-perturbative terms in the 
effective action that may be observed today. As noted earlier,
we find that these non-perturbative effects in the presence of
an ultralight particle having a mass of about $10^{-33}$ eV, give
a good fit to the observed SNe-Ia data points. The effective action of
our model has a single free parameter determined by the mass of the
particle and its curvature coupling constant. In addition, there are
 two more parameters which characterize the solutions to the effective
gravitational field equations. These can be taken to be  
 the present Hubble constant and the present 
matter density. The Hubble constant is determined by low redshift
measurements as usual. The mass of the particle, 
and the present matter density are given  by a fit to the SNe-Ia
data. 
The age of the universe, and the cosmic time, or red-shift, at which 
the non-perturbative 
contributions of the ultralight particle 
first become significant, follow from these parameters. 

%
Our model involves a renormalizable
free field theory in curved spacetime, and apart from the supposed
existence of an ultralight mass, is based on previously discovered
non-perturbative terms. Furthermore, it requires no cosmological
constant in the usual sense of the term; and no fine-tuning of 
the value of the current cosmic time is necessary to explain the 
current observed value of the matter density.

The outline of this paper is as follows.
In Sec. II, we calculate, by zeta-function regularization, the renormalized 
effective action that follows from the R-summed propagator. 
This confirms the result found with dimensional regularization 
by Parker and Toms \cite{partom}, and generalizes it to include the case when 
there is an imaginary part of the effective action. Variation of
this effective action gives the Einstein gravitational field equations,
including vacuum contributions of quantized fields. In addition, an 
imaginary part of the effective action implies a rate of particle
creation in the same way as it does in quantum electrodynamics 
\cite{schw}. The effective action has infrared-type divergences which are 
treated separately in Appendix A.

The effective action at low curvatures gives rise to induced
gravitational coupling constants, as is well-known. These constants
depend on a renormalization scale parameter. In Section III, we
consider 
the dependence of the induced
gravitational constants on the renormalization scale, and identify the
renormalized constants with their known values at low curvature. This
identification is used to fix all the coupling constants in the
effective 
action, in terms of their low curvature values,
except the field mass $m$ and its curvature coupling $\xi$.  In this
context, the conclusions of Ref. \cite{partom} are explored in greater
detail and generalized here. 
Variations of various terms in the
effective action are listed in Appendix B.

In Section IV, we clarify the meaning of exact and perturbative
solutions to the semiclassical Einstein equations. In particular, we
emphasize that a perturbative treatment in $\hbar$ must be done in
such a way  that only genuine quantum corrections are treated in a
perturbative manner, while factors of $\hbar$ which occur in the
``classical'' Klein-Gordon action are not treated
perturbatively.

In Section V, we display a set of constant curvature (de Sitter)
solutions\footnote{de Sitter solutions to the effective gravitational 
field equations 
have been found by other authors in gravity theories with higher 
derivative terms
\cite{odintsov,star}. These solutions are valid at high curvatures. In this
paper, we also find low-curvature de Sitter solutions which could play an 
important role in the late evolution of the universe.}, 
which exist even in
the absence of
an explicit cosmological constant term in the effective action. We
show that, for $\xi -1/6 > 0$, such solutions typically have
Planckian curvatures and are therefore not of much physical
interest. On the other hand, for $\xi -1/6 <0$, we show that there
exist de Sitter solutions with scalar curvature approximately equal to
$m^2/(1/6-\xi)$, and that such solutions exist for a wide range of
values of $m$ and $\xi$. As argued in later sections, these solutions
could play a vital role at the present time, if there exists an
ultralight particle with $\xi -1/6 <0$. In Section VI, we carry out
perturbative calculations of the scalar curvature about a 
classical Robertson-Walker universe,
containing classical matter and radiation. 
We find that quantum corrections to the
classical scalar curvature of the universe can become significant at an
epoch determined by the mass scale $\overline{m} = m/\sqrt{1/6 -\xi}$,
for $\xi -1/6
<0$. Furthermore, we find that the quantum corrections tend to drive the
scalar curvature to a constant value. These considerations lead to
 a scenario, presented in Section VII, 
in which quantum corrections to a matter-dominated
universe 
cause a transition to a de Sitter solution of the type discussed in
Section V. The various cosmological parameters in such a model
universe are expressed in terms of three basic parameters to be
determined by observation: the mass
scale $\overline{m}$, the present ratio of matter density to critical
density $\O_0$, and the present Hubble constant $H_0$. 

In Section VIII, we develop this scenario further, using it to obtain
magnitude-redshift curves, and comparing these curves to recent data
from high-redshift Type 1a supernovae \cite{perl}. The mass scale 
$\overline{m}$ is 
 determined to be roughly equal to
$10^{-33}$ eV, while $\O_0$ is found to be less than $0.4$. 
In Section IX, we derive from these parameters the 
age of the universe in our model, which is found to be greater than 
about $13$ Gyr.
Finally, in Section IX, we conclude with a discussion of our
results and comment on future work.    

The main results of this paper are summarized in Eqs. (\ref{ful}) and
(\ref{des}) (the
effective action of the theory in terms of measurable gravitational
coupling constants), Eq. (\ref{solut}) (the relationship
between scalar curvature and mass during the late de Sitter phase),
Eq. (\ref{modelo}) (the scale factor in our model), Eq. (\ref{lumd1})
and the equations prior to it in Section VIII (the
luminosity-distance-redshift relation), the discussion of the age 
of the universe in Section IX, and Figs. 3 and 4.

Throughout this paper, we use the metric signature convention 
$(- + + +)$, and the convention for the Riemann curvature tensor 
${R_{\m \n \rho}}^{\s} = {\G^{\s}}_{\m \rho, \n} -{\G^{\s}}_{\n \rho, \m}
+ {\G^{\a}}_{\m \rho} {\G^{\s}}_{\a \n}-{\G^{\a}}_{\n \rho}
{\G^{\s}}_{\a \m}$.  

\section{Quantum Corrections to Effective Action}

Here we derive the regularized effective
action based on the propagator for a scalar field in curved space. 

Parker and Toms \cite{partom} define the effective action via
dimensional regularization of the Feynman Green function and by
integrating the Green function with respect to the square of the
mass. Here we will define it by zeta-function regularization.

To this end, consider a $D$-dimensional scalar field theory in curved
spacetime, with classical action
\be
S= -\ha \int d^D x \,\phi(x) H(x) \phi(x),
\te
where $H(x)= -g_{\m \n }(x)\nabla^{\mu}_x\nabla^{\nu}_x +m^2 + \xi
R(x)$, and we have ignored boundary contributions.  The one-loop
effective action, $W^1$, is defined by the functional integral
\be
\label{def}
e^{iW^1} \equiv Z = \int D\phi \,e^{iS},
\te 
and is related to the transition amplitude for evolution from the
``in'' vacuum state at early times to the ``out'' vacuum state at late
times \cite{schw}:
\be
e^{iW^1} = \langle 0_{\rm out}\mid 0_{\rm in}\rangle.
\te
For free fields, as considered here, the one-loop effective action,
$W^1$, gives the {\it full} quantum contributions of the matter field
to the effective action.
By formally performing the functional integral in Eq. (\ref{def}), we
get (see, for example \cite{parcarg})
\be
\label{effo}
W^1 = \frac{i}{2}\ln {\rm det}(H/\mu^2),
\te
where we have introduced an arbitrary real quantity $\mu$ with the
dimensions of mass in order to render $Z$ dimensionless.

In terms of the zeta function, formally defined as
\be
\zeta (\nu) = {\rm Tr}\,H^{-\nu},
\te
the one-loop effective action can be expressed in the form
\be
\label{eff}
W^1 = -\frac{i}{2}[\zeta'(0) + \ln (\mu^2) \zeta(0)].
\te

The proper-time heat kernel, $K$, is defined to satisfy the equation
\be
\label{heate}
\left(i\frac{\partial}{\partial s} - H(x) \right)K(x,x',is) = 0,
\te
with initial condition
\be 
\label{ini} 
K(x,x',0) = (-g)^{-\ha}\d ^{(D)}(x-x').
\te
Because Eq. (\ref{heate}) is a Schrodinger-like equation, the heat
kernel may be formally represented as
\be
K(x,x',is) = \langle x \mid e^{-isH} \mid x'
\rangle,
\te
with the inner product $\langle x \mid x' \rangle$ defined by
Eq. (\ref{ini}).

The heat kernel gives a representation for the Feynman Green's
function of the theory:
\be
\label{gf}
G(x, x') \equiv \langle x \mid H^{-1} \mid x' \rangle = \int_0^\infty
ids \langle x \mid e^{-is(H-i\epsilon)}
\mid x' \rangle,
\te
where we have added a small imaginary part to $H$ for reasons of
convergence.

Also, using the Mellin transform to express the zeta function in terms
of the heat kernel, one obtains
\be
\label{mellin}
\zeta(\nu) = \Gamma(\nu)^{-1}\int d^Dx\sqrt{-g}\int_0^\infty ids (is)^{\nu -1}
\langle x \mid e^{-is(H-i\epsilon)}\mid x \rangle.
\te 

We now introduce the $R$-summed form of the heat kernel
\cite{partom,parja}
\be
\label{defp}
\langle x \mid e^{-is(H-i\epsilon)}\mid x' \rangle = i (4\pi i s)^
{-D/2}\D
^{\ha}(x,x')e^{i\frac{\sigma}{2s}-is(M^2-i\epsilon)}\overline{F}(x,x';is),
\te
where $\D(x,x')$ is the Van Vleck-Morette determinant and $\s (x,x')$ is
one-half the square of the geodesic distance between $x$ and $x'$. The
quantity $M^2$ is, in general, a function of $x$ and $x'$ with the
property that it reduces in the coincidence limit to
\be
M^2(x,x) = m^2 + (\xi -1/6)R(x) \equiv m^2 + \overline{\xi}R(x),
\te
where $\xib \equiv \xi -1/6$.
The form (\ref{defp}) of the heat kernel sums all terms explicitly
involving at least one factor of 
the scalar curvature $R$, in the
coincidence limit $x' \rightarrow x$. Away from the coincidence limit,
$M^2$ can be taken to be the linear combination \cite{parja}
\be
\label{lincom}
M^2(x,x') = m^2 + \xib (a R(x)+ (1-a)R(x'))
\te
 with an arbitrary choice of $a$, and the function $\overline{F}$ will
 correspondingly depend on the value of $a$.  $\overline{F} $ may be
 expanded in an asymptotic series in powers of $s$, namely,
\be
\label{sdw}
\overline{F}(x,x';is)\approx \sum_{j=0}^\infty (is)^j\overline{f}_j(x,x'),
\te
In the coincidence limit $x'\rightarrow x$, $\overline{F}$ does not
depend on $a$. We then have $\overline{f}_0(x,x)=1,
\overline{f}_1(x,x) =0$, and
\be
\overline{f}_2(x,x) = {1\over 6}\left({1\over 5}-\xi\right)\Box R +{1\over 180}
(R_{\a \b \g \d }R^{\a \b \g \d } - R_{\a \b }R^{\a \b }).
\te
The $\overline{f}_j$ for all $j$ contain no factor of $R$ (with no
derivatives acting on it).

Thus Eq. (\ref{mellin}) becomes 
\be 
\zeta (\nu) = i (4\pi)^{-D/2}\Gamma(\nu)^{-1}\int d^Dx \sqrt{-g}\intf
ids (is)^{\nu-1-D/2} e^{-is(M^2-i\epsilon)}\overline{F}(x,x;is).  
\te 
The integral over $s$ in the above expression is divergent at $\nu =0$
because of the singular behavior of the integrand as $s$ approaches
zero. This divergence actually exists for all ${\rm Re}\,\nu \leq
D/2$. [We will assume throughout this paper that $D$ is even]. We
therefore regulate the zeta function for these values of $\nu$ by
defining the zeta function for ${\rm Re}\,\nu \leq D/2$ as the
analytic continuation of the zeta function for ${\rm Re}\,\nu >
D/2$. To this end, we perform $D/2 +1$ integrations by parts before
analytically continuing, to get
\bea 
\zeta (\nu)&=& -i(-4\pi)^{-D/2}\frac{\Gamma(\nu)^{-1}}{(\nu-D/2)  
(\nu-D/2+1)\cdots \nu}\intx \intf
ids (is)^{\nu}\times \nn\\
\label{reg} & &~~~\frac{\partial^{D/2+1}}{\partial(is)^{D/2+1}}
\left(e^{-is(M^2-i\epsilon)}\overline{F}(x,x;is)\right), 
\tea 
which is regular in a neighborhood of $\nu=0$\footnote{We may equally
well analytically continue in $D$ instead of $\nu$.}. This definition
of the $\zeta$ function therefore leads to a regularised one-loop
effective action. From Eqs. (\ref{eff}) and (\ref{reg}), we get
\bea 
W^1 &=& \{2(4\pi)^{D/2}(D/2)!\}^{-1}\intx \left\{ \left(\g + \ln(\mu^2)+{1\over
D/2} + \frac{1}{D/2-1} +\cdots +1\right)\times\right.\nn \\ 
& &~~\frac{\partial^{D/2}}{\partial(is)^{D/2}}
\left(e^{-is(M^2-i\epsilon)}\overline{F}(x,x;is)\right)_{s=0} \nn \\ \label{regeff}
 & & - \left.\intf ids \ln
(is)\frac{\partial^{D/2+1}}{\partial(is)^{D/2+1}}
\left(e^{-is(M^2-i\epsilon)}\overline{F}(x,x;is)\right)\right\},
\tea
where $\gamma =$ Euler's constant $=\frac{d}{d\nu} [\Gamma(\nu
+1)]_{\nu=0}$.

Fixing the renormalization scale $\mu$ is equivalent to fixing the
(constant) phase of the out-vacuum relative to the in-vacuum in flat
space, which can be chosen arbitrarily.  We will find it convenient to
define a rescaled version of $\mu$ by $\ln (\mt^2) = \ln (\mu^2)
+{1\over D/2} + \frac{1}{D/2-1} +\cdots +1$. Then the one-loop
effective action becomes
\bea
W^1 &=& \{2(4\pi)^{D/2}(D/2)!\}^{-1}\intx \left\{ (\g +\ln(\mt^2))
\frac{\partial^{D/2}}{\partial(is)^{D/2}}
\left(e^{-is(M^2-i\epsilon)}\overline{F}(x,x;is)\right)_{s=0}\right. \nn \\
 & & ~~- \left.\intf ids \ln
(is)\frac{\partial^{D/2+1}}{\partial(is)^{D/2+1}}
\left(e^{-is(M^2-i\epsilon)}\overline{F}(x,x;is)\right)\right\}.
\tea

We may now substitute the $R$-summed Schwinger-DeWitt expansion
(\ref{sdw}) in the above formula for the effective action and perform
the required differentiation and integration term by term. In doing
so, we use the integral identity
\be
\intf ids \ln (is) (is)^p e^{-is(M^2 -i\epsilon)} = - \frac{p!}{(M^2 -i\epsilon)^{p+1}}
\left\{\ln (M^2 -i\epsilon) + \gamma -1 -\ha -\cdots -\frac{1}{p}\right\},
\te 
for integer values of $p$.  This procedure finally yields
\be
W^1 = \{2(4\pi)^{D/2}\}^{-1}\intx\left\{I_1(x) + I_2(x)+
I_3(x)\right\},
\te
where
\bea
I_1 &=&- (D/2+1)\sum_{l=D/2+1}^{\infty}\overline{f}_l\,
l!\,(M^2-i\epsilon)^{-l+D/2}
\sum_{p=0}^{D/2+1}\frac{(-1)^p}{p!(D/2+1-p)!}\sum_{n=1}
^{l+p-D/2-1}n^{-1} \\
I_2 &=&  (D/2+1)\sum_{p=1}^{D/2+1}\sum_{l=D/2+1-p}^{D/2}
\frac{(-1)^p}{p!(D/2+1-p)!}\left(\g +\ln(M^2 -i\epsilon)-\sum_{n=1}
^{l+p-D/2-1}n^{-1}\right)\times \nn \\
& &~~~~\overline{f}_l\, l!\,(M^2-i\epsilon)^{-l+D/2}. \\
I_3 &=& (\g + \ln (\mt ^2)) \sum_{p=0}^{D/2}
\frac{(-M^2)^p}{p!} \overline{f}_{D/2-p}.
\tea

In two spacetime dimensions, the above formulae lead to the one-loop 
effective action 
\bea
W^1_{(D=2)} &\approx& (8\pi)^{-1}\intxt\left\{M^2\left(\ln \mid
\frac{M^2}{\mt^2}\mid -i\pi \q (-M^2)\right)\right.\nn \\ &
&~~~\left.+\sum_{l=2}^{\infty}((l-1)^{-1}-l^{-1})
\frac{l!}{(M^2-i\epsilon)^{l-1}}\overline{f}_l \right\} .
\tea  
Similarly, in four spacetime dimensions, we get
\bea
I_1 &=&-\ha \sum_{l=3}^{\infty}\overline{f}_l \frac{l!}{(M^2
-i\epsilon)^{l-2}}\left\{ l^{-1}+2(l-1)^{-1}-(l-2)^{-1}\right\} \\ I_2
&=& -{3\over 2}\overline{f}_2 -\ha (\g +\ln \mid M^2 \mid - i\pi \q
(-M^2))(2\overline{f}_2 + M^4) \\ I_3 &=& \ha (\g +\ln (\mt ^2))(M^4
+2\overline{f}_2),
\tea
and the one-loop effective action takes the form
\bea
\label{effser}
W^1_{(D=4)} &\approx& -(64\pi^2)^{-1}\intxf\left\{(M^4
+2\overline{f}_2)\left(\ln \mid
\frac{M^2}{\mt^2} \mid 
-i\pi \q (-M^2)\right)+3\overline{f}_2 \right.\nn \\
& &~~~\left.+ \sum_{l=3}^{\infty}(l^{-1}+2(l-1)^{-1}-(l-2)^{-1})
\frac{l!}{(M^2-i\epsilon)^{l-2}}\overline{f}_l \right\} .
\tea

This is the $R$-summed form of the effective action.
The Gaussian approximation   amounts to keeping only the term 
multiplying $M^4$ in the above series. The terms involving $\overline{f}_2$ are
required
to obtain the 
correct trace anomaly.

The first two terms in curly brackets in Eq. (\ref{effser}) agree with
the earlier dimensional regularization results of Parker and Toms,
except for the step function term $\q (-M^2)$ which is purely
imaginary and implies particle production when $M^2$ becomes negative
(Parker and Toms assumed $M^2>0$). This step function term originates
from the identity $\ln (x -i\epsilon)= \ln\mid x \mid -i\pi\q (-x)$.
Within the $R$-summed form, therefore, the particle production
takes on a very simple form, with $M^2 =0$ being the threshold for
vacuum instability, or the creation of particle pairs. It is 
conceivable that there are physical situations where the imaginary
part of the first two terms in the above formula for the effective
action very closely approximates the actual gravitational particle
creation. In this paper, we do not deal with this issue further since
it does not affect our results.

Note that Eq. (\ref{effser}) is only an asymptotic series expansion in
inverse powers of $M^2$, arising out of an expansion of the heat
kernel which ignores, for example, terms that have essential
singularities at $s=0$. 
Only the first two terms (i.e. up to $\overline{f}_2$)
are necessary for renormalization and for the correct trace
anomaly. Also, these terms include the convergent infinite sum
involving the scalar curvature. Therefore, the approximate effective
action based on these first two terms is sufficient to indicate the
non-perturbative effects coming from the infinite sum of scalar
curvature terms. This is the form of the effective action that we employ.

\section{Renormalization and Observable Gravitational Couplings}

Although we are dealing with a free field theory here, the logarithmic
dependence of the effective action on the curvature in
Eq. (\ref{effser}) leads to non-trivial effects in strong curvature
regions \cite{partom}.  Before we go on to a discussion of such
effects, it is necessary to understand the contribution of the one-loop
effective action to the full gravitational action in regions of low
curvature. Also, we will now consider the possibility of having
multiple particle species contributing to the one-loop effective
action. For particles of spin $\ha$ and $1$, it has been shown in
Ref. \cite{parja} that all terms involving $R$ in the heat kernel can
be summed in a similar manner to the spin 0 case, in a simple
exponential form. We will therefore consider a generalization of
Eq. (\ref{effser}) to the form
\be
\label{mult}
W^1 = -\hbar (64\pi^2)^{-1} \sum_i n_i \intxf \left\{(M_i^4 + p_i
\overline{f}_{2i})\ln \mid M_i^2/\mt _i^2 \mid + q_i \overline{f}_{2i}
+ \ldots \right\},
\te
where we have inserted arbitrary, species-dependent coefficients
$n_i$, $p_i$ and $q_i$, and a factor of $\hbar$ is now made
explicit. 
The imaginary term is not shown because we are interested here in the
part of the effective action that corresponds to vacuum
polarization. In the low curvature limit considered in this section,
we can assume that $M_i^2 = m_i^2 + \xib R >0$, so there is no
imaginary term. In later sections where $M_i^2$ can be negative but
the curvature is very small with respect to the Planck scale, the
created particles make a negligible contribution to the classical
matter already present in the Robertson-Walker universe under
consideration. 
The terms of higher
order in the curvature correspond to the asymptotic series sum in
Eq. (\ref{effser}) and do not appear in Eq. (\ref{mult}). 
For higher spin fields, the heat kernel (and the
Green's function) generally has a matrix structure. In evaluating the
effective action in such cases, one performs an additional trace over
all internal indices. It will be understood that such a trace has been
carried out before arriving at Eq.  (\ref{effser}). $\overline{f}_2$
will be understood to mean the trace of the modified second
Schwinger-DeWitt coefficient. The values of species-dependent
coefficients for spin $1/2$ are $\xib =
1/12$, $n_i = -4$, $p_i = 1/2$, and $q_i = 3/2$.

At low scalar curvature ($\xib R \ll m^2$), one may expand the
logarithm in Eq.  (\ref{mult}) in powers of $R$. Noting that
$\overline{f}_{2i}$ can be expressed as the linear combination
\be
\overline{f}_{2i} = a_i \Box R + b_i R_{\a \b }R^{\a \b } + 
c_i R_{\a \b \g \d }R^{\a \b \g \d },
\te
the leading terms in the 1-loop effective action then give
\bea
\label{one}
W^1 &\simeq &-\hbar(64\pi^2)^{-1}\sum_i
n_i\intxf\left\{\vphantom{\ha}m_i^4\ln(m_i^2/\mt _i^2) + m_i^2\xib_i
R\left(1+2\ln(m_i^2/\mt_i^2)\right)\right. \nn\\ & &\left.~+ \xib_i^2
R^2 \left(3/2 + \ln (m_i^2/\mt_i^2)\right) + \left(b_i R_{\a
\b } R^{\a \b } + c_i R_{\a \b \g \d } R^{\a \b \g \d } \right)
(p_i \ln (m_i^2/\mt_i^2) + q_i)+ \ldots
\right \},
\tea
In the above expression, we have allowed for different renormalization
points characterized by the different mass scales $\mt _i$. Changing
these mass scales give rise to terms that can be absorbed into the
bare gravitational action, as will be discussed below.  However, one
can still remove the $\mt_i$-dependence of the full effective action
by using our {\it knowledge} of the observed gravitational coupling
constants. To be precise, consider the bare gravitational action
\bea
\label{bare}
W_g &=& \intxf \left\{(\k +\d \k )(R -2(\L +\d \L))+ (\a_1 +\d
\a_1)R^2 +(\a_2 +\d \a_2 )R_{\m \n }R^{\m
\n } \right.\nn \\
& &~~~~\left.+ (\a_3 + \d \a_3)R_{\m \n \g \d }R^{\m \n \g \d }\right\}
\tea 
where the counterterms are
\bea
\label{counter}
\d \L &=& -(128\pi^2\k)^{-1} \hbar\sum_i n_i m_i^4 \ln(\m _1^2/\mt_i^2) \nn \\
\d \k &=& (32\pi^2)^{-1} \hbar\sum_i n_i m_i^2 \xib_i\ln(\m _2^2/\mt_i^2)\nn \\
\d \a_1 &=& (64\pi^2)^{-1} \hbar \sum_i n_i \xib_i^2
\ln(\m _3^2/\mt_i^2) \nn \\
\d \a_2 &=& (64\pi^2)^{-1} \hbar \sum_i n_i b_i p_i
\ln(\m _4^2/\mt_i^2) \nn \\
\d \a_3 &=& (64\pi^2)^{-1} \hbar \sum_i n_i c_i p_i
\ln(\m _5^2/\mt_i^2).
\tea
$\m _{1,2,3,4,5}$ are arbitrary constants of dimension mass, and the
$\mt_i$-dependence of the counterterms is required, following
\cite{partom}, to cancel the $\mt_i$-dependence in the 1-loop
effective action\footnote{In zeta function regularization, the
divergent pieces of the one-loop effective action have been thrown
away beforehand, which is why it is not necessary to introduce those
divergences into the counterterms either (although this procedure of
introducing counterterms could be bypassed in zeta-function
regularization, it is necessary in other regularization schemes, such
as dimensional regularization). In dimensional regularization
\cite{partom}, one explicitly keeps track of the divergent pieces and
introduces corresponding divergences in the counterterms. Since the
divergences are ultimately canceled, one finally ends up with
Eq. (\ref{obs}) in any case.}.  Adding Eqs. (\ref{one}) and
(\ref{bare}) we obtain the full effective action at low curvatures, as
\be 
\label{fulll}
W \equiv W_g +W_1 = \intxf \left\{-2\k_o \L_{o} + \kappa_{o} R + \a
_{1o}R^2 + \a _{2o}R_{\a \b }R^{\a \b } + \a _{3o}R_{\a \b \g \d
}R^{\a \b \g \d } + \ldots\right\},
\te
where the subscipt o refers to the observed gravitational constants at
low curvatures. These are combinations of the bare and induced
constants, independent of $\mt_i$, and given by
\bea
\label{obs}
-2\k_o \L _{o} &=& \L - \hbar(64\pi^2)^{-1}\sum_i n_i m_i^4
\ln(m_i^2/\m _1^2) \nn \\
\k _{o} &=& \k - \hbar (32\pi^2)^{-1}\sum_i n_i m_i^2 \xib_i 
\left(\ln(m_i^2/\m _2^2)+\ha \right)\nn \\
\a _{1o} &=& \a_1 - \hbar (64\pi^2)^{-1}\sum_i n_i \xib_i^2 
\left(\ln (m_i^2/\m _3^2)+ {3\over 2}
\right)\nn \\
\a _{2o} &=& \a_2 - \hbar (64\pi^2)^{-1}\sum_i n_i b_i 
\left(p_i\ln (m_i^2/\m _4^2)+ q_i \right)\nn\\
\a _{3o} &=& \a_3 - \hbar (64\pi^2)^{-1}\sum_i n_i c_i 
\left(p_i\ln (m_i^2/\m _5^2)+ q_i \right)
\tea
Note that one can always absorb the dependence of the observed
constants on the $\m _i$'s into the bare constants. We thus have some
freedom in shifting terms within the above equations.  However, as
already stated, we may, in principle, use our knowledge of $\L _{o}$,
$\k _{o}$ etc. in Eq. (\ref{fulll}) at {\it low curvature} to explore
the theory in regions of high curvature. Indeed, the full effective
action in regions of high curvature now depends only on the observed
values of the gravitational coupling constants at {\it low curvature},
on the physical particle masses $m_i$ and on the values of $\xib_i$
(these are fixed for higher spin fields). No other parameters enter
into the effective action.  To see this, we add Eqs. (\ref{mult}) and
(\ref{bare}), use Eq. (\ref{counter}) to substitute for the
counterterms, and finally use Eq. (\ref{obs}) to eliminate the five
mass scales in favor of the observed constants. In this procedure, the
$\mt_i$ dependence in Eq. (\ref{mult}) and the $\mu_i$ dependence in
Eq.  (\ref{obs}) cancel the corresponding dependences in the
counterterms of Eq. (\ref{counter}). Then the full effective
action below threshold is given by
\bea
\label{ful}
W \equiv W_g + W^1 &=& \intxf \left\{-2\k_o \L _{o} - \hbar
\frac{1}{64\pi^2}\sum_i n_i m_i^4 \ln \mid
\frac{M_i^2}{m_i^2}
\mid \right.\nn \\
& &+\left(\k _{o} + \hbar \frac{1}{64\pi^2}\sum_i n_i m_i^2 \xib_i
\left(1-2\ln \mid \frac{M_i^2}{m_i^2}
\mid \right)
\right) R\nn \\
& &-\hbar\frac{1}{64\pi^2}\sum_i n_i p_i\ln \mid
\frac{M_i^2}{m_i^2}\mid f_{2i} +
\left(\a _{1o}+ \hbar \frac{3}{128\pi^2}\sum_i n_i \xib_i^2 \right) R^2
\nn \\
& & +\left. \a_{2o} R_{\a \b }R^{\a \b } + \a_{3o} R_{\a \b \g \d
}R^{\a \b \g \d } + \ldots \right \},
\tea
where $f_{2i}$ is given by
\be
p_i f_{2i} = p_i \overline{f}_{2i} + \ha \xib_i^2 R^2.
\te
Here, 
\be
\k_o = (16\pi G_o)^{-1},
\te
where $G_o$ is Newton's gravitational constant, and $\L_o$ is the
usual cosmological constant.

Eq. (\ref{ful}) is a new result, relating behavior at high curvatures
to values of the gravitational coupling constants at low
curvatures. For scalar fields, $\xib$ is the only undetermined
parameter in the effective action. For higher spins, the corresponding 
parameters have fixed values.  We emphasize again that terms involving 
$\overline{f}_3$ and higher,
omitted in Eq. (\ref{ful}), constitute an asymptotic expansion in 
inverse powers of
$M^2$ and are therefore not necessarily expected to be physically
significant. The terms retained in Eq. (\ref{ful}) are the minimal set
of terms necessary for renormalization and incorporate the
 sum of scalar curvature terms in the propagator.
It is readily checked that
Eq. (\ref{fulll}) constitutes the first few terms in the
low-scalar-curvature limit ($\mid \xib R \mid \ll m^2$) of
Eq. (\ref{ful}). 

In subsequent Sections, we consider the gravitational field equations
for a single scalar field, as derived from Eq. (\ref{ful}), 
and their consequences. Comparison of Eqs. (\ref{mult}) and
(\ref{effser}) gives, for scalar fields, $n_i =1, p_i =2, q_i =3$. 
In Section V, we
study the case of constant curvature spacetimes. In later sections, we
generalize the investigation to Robertson-Walker universes containing
matter and radiation. In the next Section, we clarify the usage of the
terms ``perturbative'' and ``exact'' in this paper.

\section{Meaning of Exact and Perturbative Solutions}

In the previous section we derived the effective action,
Eq. (\ref{ful}), appropriate to the $R$-summed propagator. 
All terms multiplied by $\hbar$ in Eq. (\ref{ful}) will be
called ``quantum-vacuum'' terms, also referred to as one-loop terms,
because they arise after vacuum fluctuations of the field have been
integrated out in a path-integral formulation of the theory.  However,
note that factors of $\hbar$ must also be included in the argument of
the logarithmic terms for dimensional reasons. Indeed, one has
\be
\ln \mid M^2/m^2 \mid = \ln \mid (m^2 + \hbar^2 \xib R)/m^2 \mid.
\te
An approach that is completely perturbative in $\hbar$ would then
involve expanding out the logarithmic terms.  However, the factor of
$\hbar^2$ appearing in the above equation has its origins in the
generalized Klein-Gordon  equation
\be
\left(-\Box + \frac{m^2}{\hbar^2} +\xi R \right) \phi =0,
\te
whose solutions contribute to the {\it tree-level} effective action,
as opposed to the quantum-vacuum or one-loop effective action. That
is, the $\hbar^2$ factor above does not arise from integrating out
vacuum fluctuations. Furthermore, the quantity $m^2/\hbar^2$ is the
square of the inverse Compton wavelength of the field, and need not be
large relative to the curvature scalar. Therefore, it is reasonable to
regard the logarithmic terms as non-perturbative when expanding in
$\hbar$. It is convenient to avoid explicitly inserting factors
of $\hbar$ in the arguments of logarithmic terms (more generally, in
any term which contains $M^2$), with the understanding that the mass
of the field is interpreted as an inverse Compton wavelength.

A {\it perturbative} analysis in $\hbar$, as used in Section VI of
this paper, therefore treats only the quantum-vacuum corrections in a
perturbative fashion, while keeping the logarithmic terms intact. It
is not inconsistent to do so, since it defines a regime in which the
semiclassical corrections to the effective action are much smaller
than the tree-level terms involving $\L_o$ and $\k_o$, yet in which
the scalar curvature is allowed to be of the same order as the square
of the inverse Compton wavelength. On the other hand, an
{\it exact} analysis, as used in Section V below, treats even the
quantum  corrections in an exact fashion.

The reason we carry out both perturbative and exact
analyses is as follows. Exact solutions play a significant
role when the perturbative analysis breaks down
(as signaled by a rapid growth in the contributions of quantum
corrections to the metric). In order to obtain an understanding of the
full dynamics of the metric, it is therefore necessary to construct
both perturbative and exact solutions to the semiclassical
Einstein equations.

It is well-known \cite{simon} that semiclassical equations of motion
can be perturbatively reduced in a manner such that the resulting
equations only admit solutions that are perturbative in $\hbar$. It
has been further argued that such solutions are the only physically
viable solutions of the full semiclassical equations since they do not
exhibit runaway behavior in the classical limit.

However, in Section V, we argue for the physical significance of
our exact solutions. This argument is based on two
observations. First, owing to the presence of a mass scale in the
theory, $\hbar$ explicitly enters into the semiclassical equations
only via the dimensionless ratio $r \equiv m^2/m_{Pl}^2$.  This ratio 
is not necessarily small, although perturbation theory
assumes that it is. Thus, there could be physical effects at large
$r$, not encountered in perturbation theory.

Secondly, we will find exact solutions that cannot be expanded in
$\hbar$ and have a
well-defined limit as $r \rightarrow 0$, i.e.  they do not possess
runaway behavior in the classical limit. Such solutions must therefore
be regarded as physical solutions.

For these reasons, we believe that exact solutions that do not arise
from perturbative reduction must be
included in a complete analysis of solutions of the semiclassical
Einstein equations, at least when a mass scale is present in the
theory.

\section{Exact Vacuum de Sitter Solutions}

In this section, we will consider exact solutions to the
equations of motion specialized to de Sitter spacetime. These
equations simplify considerably for spacetimes of constant curvature.
This simplification allows us to include in the effective action terms
involving $\overline{f}_2$ and $R^2$. 

As we shall see, there is a rich variety of constant curvature de
Sitter solutions, with the scalar curvature being highly sensitive to
the value of $\xib$. We will argue that these solutions are physically
viable although they do not appear in the perturbative reduction
approach.

Consider, therefore, the effective action of Eq. (\ref{ful}),
specialized to a single scalar field with mass $m$ and curvature
coupling $\xi$:
\bea
\label{des}
W &=& \intxf \left\{-2\k_o \L _{o} - \hbar \frac{1}{64\pi^2} m^4 \ln \mid
\frac{M^2}{m^2}
\mid \right. \nn \\
& &+\left.\left(\k _{o} + \hbar \frac{1}{64\pi^2} m^2  
\xib \left(1-2\ln \mid \frac{M^2}{m^2}
\mid \right)
\right) R -\hbar\frac{1}{32\pi^2}\ln \mid \frac{M^2}{m^2}\mid f_{2} +
\hbar \frac{3}{128\pi^2}\xib^2  R^2 \right\},
\tea
where we have assumed that the constants $\a_{1o}$, $\a_{2o}$ and
$\a_{3o}$ are negligibly small\footnote{For the purposes of this
section, it is actually sufficient to assume that terms involving
$\a_{1o}$, $\a_{2o}$ and $\a_{3o}$ combine to yield
$\a_{o}f_2+$negligible contributions, where $\a_{o}$ is some constant,
because the variation of $\intxf f_2$ vanishes in a constant curvature
spacetime.}.

One may now use Eqs. (B1-B9) of Appendix B to obtain the equations of
motion resulting from the variation of Eq. (\ref{des}). When
specialized to a constant curvature maximally symmetric spacetime, 
in which
\be
R_{\m \n \a \b } = \frac{R}{12}(g_{\m \a}g_{\n \b}-g_{\n \a}g_{\m \b}),
\te
these equations yield a
single algebraic equation satisfied by the scalar curvature
$R$:\footnote{The right-hand-side (RHS) of Eq. (\ref{alg}) gives the 
trace of a 
stress-tensor in de Sitter space
which is not the same as the stress-tensor for the Bunch-Davies
vacuum state. The stress-tensor on the RHS of Eq. (\ref{alg})
corresponds to a different state 
and is defined by variation of $W$. 
The leading terms of the stress-tensor agree with those arising 
from the Gaussian
approximation of Bekenstein and Parker \cite{bekpa}, which is known to
be a good approximation  in de Sitter space, particularly for closely
spaced points as are used in defining the stress tensor.
The additional term, $(1/2)(1/1080)\xib (R^3/M^2)$ in Eq. (\ref{alg})
yields 
the correct
trace anomaly. 
Furthermore, this stress-tensor is conserved, and
vanishes in flat spacetime ($R=0$). The RHS of Eq. (\ref{alg})
diverges at $M^2 = m^2 + \xib R = 0$. Similar divergences also occur
in
 the stress tensor in the
Bunch-Davies vacuum state in de Sitter space, which diverges  
at $m^2 + (\xi+n(n+3)/12)R =0$, where $n$ is a
non-negative integer.}
\bea
\label{alg}
2\k_o (4\L_o-R) &=& -\frac{\hbar}{16\pi^2}\left\{
m^4 \ln \mid \frac{M^2}{m^2} \mid \left(1+ \xib \frac{R}{m^2}
\right) - \ha m^2 \xib R \left(1 +\frac{m^2}{M^2}\right)\right.\nn \\
& & \left.- \xib^2 R^2 \frac{m^2}{M^2} - \ha \xib
\frac{R^3}{M^2} \left(\xib^2 - {1\over 1080}\right)\right\}.
\tea
Note that we recover Starobinsky inflation by first taking $\L_o
\rightarrow 0$ and $m
\rightarrow 0$, and then taking the limit $\xib
\rightarrow 0$. The resulting solution has scalar curvature $R= 69120\, 
\pi^2 \k_o \hbar^{-1}$, in agreement with
Starobinsky's results \cite{star}. The Starobinsky solution is
non-analytic in $\hbar$ and is not well-defined in the limit $\hbar
\rightarrow 0$. The existence of such solutions has motivated arguments in
favor of the perturbative reduction scheme \cite{simon}, which
discards such solutions in a self-consistent manner.

However, such arguments weaken in the presence of an additional mass
scale $m$ in the theory.  To see this, consider Eq. (\ref{alg})
rewritten in terms of the dimensionless variables
\bea
y &=& \frac{R}{m^2} \\
r &=& \frac{\hbar m^2}{16\pi \k_o}.
\tea
Recall that $m$ refers to the inverse Compton wavelength of the
particle, so $y$ and $r$ are indeed dimensionless. In terms of the
actual mass of the particle, $r$ is given by 
\be
r = \frac{m_{{\rm actual}}^2}{m_{Pl}^2},
\te
where $m_{Pl} = \sqrt{16 \pi \k_o \hbar}$.

Eq. (\ref{alg}) then takes the form
\be
\label{dimalg}
(1+\xib y) \ln \mid 1+\xib y \mid - \frac{\xib y}{1+\xib y} \left\{ 1
+{3\over 2}\xib y +\ha y^2
\left(\xib^2 - {1\over 1080}\right) \right\} = \frac{2\pi}{r}
\left(y- \frac{4\L_o}{m^2}\right).
\te
The solution for $y$ is a function of three dimensionless parameters:
$r$, $\xib$ and $\L_o/m^2$. In a perturbative reduction approach, $r$ is
regarded as a small parameter and the solution for $y$ is constrained
to be an analytic function of $r$. However, it is plausible that the
limit $\hbar \rightarrow 0$ does not imply $r
\rightarrow 0$, i.e. that the mass $m$ is rescaled in such a manner 
that $r$ stays constant (or roughly constant)
as $\hbar \rightarrow 0$. For example, in string theory, one would
expect that all masses (including the Planck mass $m_{Pl}$) are
generated by a single scale (the string scale), in which case the
dimensionless quantity $r$ would be independent of $\hbar$. In the
absence of knowledge of the precise nature of a fundamental unified
theory, it is therefore prudent to consider the possibility that the
parameter $r$ is some finite quantity, not necessarily small, and to
treat it in a non-perturbative fashion\footnote{Different arguments
for the physical significance of  solutions that do not arise from
perturbative reduction
have been given by Wai-Mo Suen \cite{suen}.}.  In subsection B, we
give a further justification for the physical validity of
solutions that arise without expanding in $r$. Namely, we find that 
for $\xib <0$, these
solutions have a well-defined classical limit as $\hbar$ (or $r$)
$\rightarrow 0$.

We now consider numerical solutions to the algebraic equation
(\ref{dimalg}). This equation, in general, has solutions with both
positive and negative scalar curvature. We will, however, focus on the
positive scalar curvature solutions ($y>0$), corresponding to an
inflating de Sitter universe. Figs. 1 and 2 are plots of the left hand side
(LHS) and right hand side (RHS) of Equation (\ref{dimalg}), as
functions of $y$, for various values of the three parameters $r$,
$\xib$ and $\L_o/m^2$. The point(s) of intersection of the LHS and RHS
correspond to solutions for $y$. These plots are convenient ways of
identifying the solution space because the LHS depends solely on the
parameter $\xib$, while the RHS is a linear function of $y$, with
slope given by $r$, and intercept given by $\L_o/m^2$. In all plots,
therefore, the straight line is the RHS of
Eq. (\ref{dimalg}). Increasing the value of $r$ will decrease the
slope of the straight line. Increasing $\L_o$ will shift the straight
line up. It is convenient to consider two ranges of values of $\xib$ that give
qualitatively different behavior of the LHS of
Eq. (\ref{dimalg}). These are: a) $\xib > 0$, and b) $\xib <0$.

We will find that there exist solutions with non-zero scalar
curvature, even if $\L_o =0$. For $\xib > 0$, the most interesting
solutions of this type exist for values of $\xib$ very close to
$(1080)^{-\ha}$. For other values of $\xib >0$, there are either no
solutions or solutions for which the scalar curvature is of order
$m_{Pl}^2$, which may thus be unphysical.

For $\xib<0$, we find that there exist solutions with $R \simeq
-m^2/\xib$, for a large range of values of $\xib$ and $r$, and for
small and large values of $\L_o/m^2$. These solutions are of greater
interest for the purposes of this paper, and the reader may safely
skip directly to subsection B.
 
\subsection{Solutions with $\xib >0$}

For $\xib > 1080^{-1/2}$ and $\L_o =0$, the 
only solution to
Eq.
(\ref{dimalg}) is the trivial solution $y=0$, because, for $y>0$, the
LHS is always negative, while the RHS is always positive. However,
addition 
of a non-zero value of $\L_o$ allows the RHS to take negative values,
 and leads to a non-trivial solution
with a value of $y$ slightly lower than the classical value
$4\L_o/m^2$. The deviation from classical behavior will typically be
very small due to the extreme flatness of the LHS graph near the
origin.

\begin{figure}[hbt]
\centering
\leavevmode
\epsfysize=5.in\epsffile{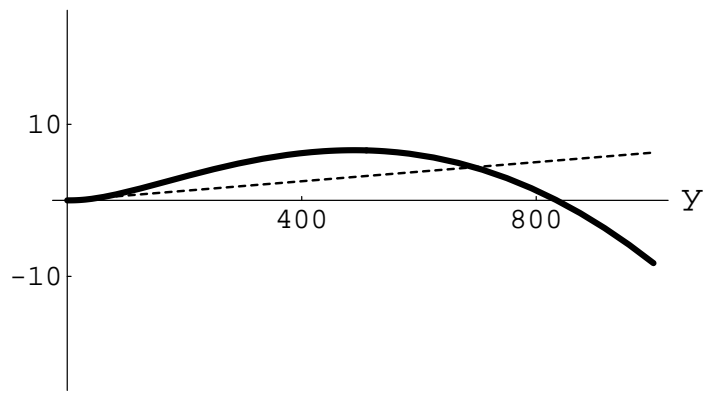}
\caption{A plot of the LHS (bold-faced curve) and RHS (dashed line)
of Eq. (49),  as
functions of
$y$, for $\xib=0.033$, $r=10$ and $\L_o=0$. The slope of the dashed
line increases as $r$ decreases.}
\label{figu3}
\end{figure}

As one lowers the value of $\xib$ towards $(1080)^{-1/2}$, the LHS
 acquires a local maximum (Fig. \ref{figu3}).  The position of
this maximum is very sensitive to the value of $\xib$ when $\xib$ is
close to the value $(1080)^{-1/2}$. For large values of $r$ ($r=10$ in
the figure), a non-trivial intersection with the RHS graph is now
possible, with $y$-values ranging from about a hundred to extremely
large values, depending on the precise values of $\xib$ and $r$. The
presence of a non-zero $\L_o$ term would shift the RHS graph down and
increase the value of $y$ even further.

When $\xib = (1080)^{-1/2} = 0.03042\ldots$, the non-trivial
intersection point of the two graphs will occur at an extremely large
value of $y$, for typical values of $r$. One may obtain an analytical
estimate by considering an approximation to Eq. (\ref{dimalg}) for
large $y$, after setting $\xib = (1080)^{-\ha}$. This gives
\be
\ln \frac{y}{\sqrt{1080}} = \frac{2\pi}{r}\sqrt{1080} + {3\over 2}.
\te
For $r=1$ ($m=m_{Pl}$), the above equation implies a value of $y$
approximately equal to about $10^{91}$.  This value {\it increases} as
$r$ {\it decreases} (i.e. by decreasing $m$ and holding $m_{Pl}$
constant) and vice-versa. However $R=m^2y = m_{pl}^2 r y$ will acquire
a minimum value for some $r>1$. Also, addition of a $\L_o$ term tends
to decrease the value of $y$ and, consequently, $R$.

For $0<\xib<(1080)^{-1/2}$,  
there is a non-trivial solution
with an  extremely large value
of $y$. An analytic approximation to
Eq. (\ref{dimalg}) can be made once more, after noting that the term
involving $y^2$ in the LHS is now the dominant term (this term
vanishes when $\xib = (1080)^{-1/2}$). Thus we obtain, in this regime,
with zero cosmological constant,
\be
y= \frac{4\pi}{r}\left({1\over 1080}-\xib^2\right)^{-1}.
\te
$y$ therefore scales linearly with $r^{-1}$. However, the scalar
curvature itself is essentially independent of $m$ in this regime, and
is given by
\be
R= 4\pi m_{Pl}^2 \left({1\over 1080}-\xib^2\right)^{-1}.
\te
The two equations above predict that $y$ (or $R$) will decrease as
$\xib$ decreases. This behavior is borne out by numerical
study. However, when $\xib$ is extremely small, $\xib y$ can become
small, and the large-$\xib y$ approximation breaks down. For very
small values of $\xib$, the solution for $y$ then increases with
decreasing $\xib$, giving $y \rightarrow \infty$ as $\xib \rightarrow
0^{+}$, as expected from Eq.  (\ref{dimalg}).

To summarize, for $\xib >0$ and $\L_o=0$, non-zero solutions for
constant scalar curvature occur for $\xib$ values very close to
$(1080)^{-1/2}$ (this corresponds to the curvature coupling $\xi$
being very close to the conformal fixed point $\xi=1/6$, because
$(1080)^{-\ha}$ is a small number). Such solutions also occur for
$\xib=(1080)^{-1/2}$; however, in this case, the curvature is
typically many orders of magnitude greater than Planck size, and may
thus be unphysical.

\subsection{Solutions with $\xib<0$}

For $\xib<0$, the LHS of Eq. (\ref{dimalg}) becomes singular at $y =
-\xib^{-1}$. This value of $y$ becomes an exact solution to
Eq. (\ref{dimalg}) in the ``classical'' limit $r \rightarrow 0$, as
can be seen by letting $y = -\xib^{-1} + \e (r)$ in
Eq. (\ref{dimalg}), and showing that $\e (r) \rightarrow 0$ as $r
\rightarrow 0$. This
fact constitutes another argument against the perturbative reduction
scheme in this case, because the scalar curvature does not exhibit runaway behavior
in the classical limit. As we shall see, most solutions with $r<1$,
will lie very close to the value $y=-\xib^{-1}$. Significant
deviations from this value will occur only for very large values of
$r$, or when a non-zero $\L_o$ term is present.

\begin{figure}[hbt]
\centering
\leavevmode
\epsfysize=5.in\epsffile{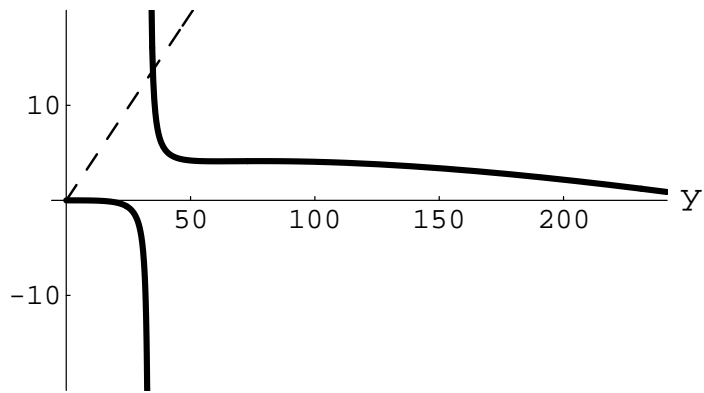}
\caption{A plot of the LHS (bold-faced curve) and RHS (dashed line) of 
Eq. (49), as
functions of
$y$, with $\xib =-0.03$, $r=16$ and $\L_o=0$. As $r$ decreases, the
slope of the dashed line increases, and the intersection point is
shifted closer to the value $y = -\xib^{-1}$. Recall that $r =
m^2/m_{Pl}^2$ and $y = R/m^2$.}
\label{figu7}
\end{figure}

Fig. \ref{figu7} is a plot of the LHS and RHS of Eq. (\ref{dimalg})
with representative values
$\xib=-0.03$ and $\L_o=0$.  An exaggerated value of $m = 4 m_{Pl}$ is
used to fully display the graph. However, as we will see, the solution
for $y$ is
largely insensitive to the precise values of $\xib$ and $r$. 

The straight line LHS graph
intersects the RHS graph at a $y$-value very slightly larger than
$-\xib^{-1}$ ($= 33.3\ldots$ in this case).  
The corresponding scalar curvature is given by $R \simeq
m^2 \xib^{-1}$. The steep vertical ascent of the LHS graph near the
value 
$y=-\xib^{-1}$
implies that the solution for $y$ is not very sensitive to the precise
value of $r$ (note that decreasing $r$ increases the slope of  the
straight line, giving a solution even closer to the value $\xib^{-1}$).
 For most values of $r$, except for exceptionally large values,
we  have a solution at $R = -m^2 \xib^{-1}$ plus a small
correction. 
The
presence of a $\L_o$ term  has the effect of shifting down the RHS
graph, leading to larger values of the scalar curvature. However, in
this case, a second solution will appear with $y<-\xib^{-1}$, the $y$-value
being very close to the classically expected value $4\L_o/m^2$, for a
small cosmological constant.  As the cosmological constant is
increased to large values this second solution will now approach $y=-\xib^{-1}$,
while the first solution now corresponds to the
classically expected large value of $y$. The solution $y\simeq-\xib^{-1}$ is
therefore fairly robust; it exists for large ranges of the parameter
$r$ and for both small and large values of $\L_o$. Numerical
investigation shows that such a
solution also exists for a large range
of values of $\xib$.

It should be noted that a $y$ value slightly larger than $-\xib^{-1}$
 corresponds to a small negative value of $M^2$, and therefore the
 effective action evaluated on such a solution acquires a small
 imaginary part giving rise to a small rate of particle production. 
In the regime we consider in this paper, the effects of this particle
production are negligible.
  
The qualitative behavior of the LHS of Eq. (\ref{dimalg}) for $\xib =
-(1080)^{-\ha}$ is very similar to the behavior for $\xib=-1$, and
quite insensitive to the precise value of $r$, as in the previous
case. The solution will now be very close to $y = -\xib^{-1} \simeq
32.86\ldots$, implying that the scalar curvature is about an order of
magnitude larger than $m^2$, for a large range of values of
$m$. Again, this is a robust solution as discussed earlier.

To summarize, for $\xib<0$, we find physically reasonable values (in
the sense that they could be small with respect to $m_{Pl}^2$) of the
scalar curvature, approximately given by
\be
\label{solut}
R \simeq -\frac{m^2}{\xib} = \overline{m}^2
\te
for a large range of values of $r$, $\xib$ and $\L_o$. The
approximation in Eq. (\ref{solut}) breaks down if $\L_o=0$ and $r \gg
1$, in which case the only solution corresponds to an extremely large
scalar curvature ($R \gg m_{Pl}^2$) whose precise value depends on $r$
and $\xib$. This large $R$ solution will give rise to a large
imaginary contribution to the effective action, and so one expects
copious amounts of particle production to occur, which could, in turn,
bring the scalar curvature down to reasonable values. This would be
consistent with a gravitational Lenz's law mechanism
\cite{parker,parker2,parrav}. We do not
address this issue here. Instead, we now turn to a 
perturbative analysis of the semiclassical Einstein equations.

\section{The growth of quantum corrections to the scalar curvature}

In this section, we will analyze,  the
effect of  logarithmic curvature terms in the effective action on a Robertson-Walker
cosmology. We analyze this effect in two ways, valid for spatially
open, closed and flat models. First, we consider a
universe with mixed matter and radiation and assume that the scalar
curvature is slowly varying so that its derivatives can be
ignored. This analysis shows that perturbative (as defined in Sec. IV)
quantum corrections to the scalar
curvature can sharply increase in magnitude near a time $t_j$
that is determined by $m$ and $\xib$. In particular, the existence of
an ultralight mass in the theory can lead to significant quantum
effects close to the present time. Since the background classical
scalar curvature is decreasing, the effect of quantum corrections is
to prevent the scalar curvature from decreasing further. However, as
quantum 
corrections become
more and more significant, perturbation theory breaks down and  cannot
be relied upon to give the full behavior of the scalar
curvature. We then carry out  a second
analysis of the behavior of the scalar curvature for all times after $t_j$,
without using perturbation theory. This second analysis indeed reveals
that, for $t>t_j$,
 the scalar curvature tends to decrease extremely
slowly, ultimately approaching a
constant value. 
The analysis thus displays consistency with the original assumption of
slowly varying scalar curvature.

In Section VII, we will use the behavior of the scalar curvature to construct a 
model universe in which a matter-dominated cosmology transits to a
mildly inflating de
Sitter cosmology at the time $t_j$.

\subsection{Quantum Corrections to the scalar curvature}
In this subsection, we will consider a classical Robertson-Walker
cosmology with mixed matter and radiation, and treat the quantum
effects involving logarithmic curvature terms in a perturbative
fashion (recall the discussion in Section IV, which shows why an
expansion of the logarithm itself is not appropriate in a perturbative
treatment of quantum-vacuum terms). The universe will deviate from classical
behavior when the quantum effects become sufficiently large. The
essential idea is to allow for the possibility of  quantum effects
being significant at the present time. We will find that this is
possible if there exist very light mass fields with $\xib<0$.

Our starting point for the analysis here is the effective action of
Eq. (\ref{des}). Variation of this effective action, specialized to a
Robertson-Walker metric, will generally yield terms involving
derivatives of the scalar curvature $R$, in addition to the terms of
Eq. (\ref{alg}). In this subsection, we  assume that terms
involving derivatives of the scalar curvature are negligible. We will
find that, for light mass fields ($m \ll m_{Pl}$), this assumption is
justified because derivatives of the scalar curvature remain small
until the magnitude of the quantum-vacuum contribution to the scalar
curvature itself becomes comparable to the classical contribution to
the scalar curvature. Beyond this point, the perturbative analysis
breaks down. In the next subsection, we carry out an exact analysis,
valid for $t>t_j$.

Although we ignore derivatives of the scalar curvature when carrying
out the variation of Eq. (\ref{des}), the resulting semiclassical
Einstein equations differ from Eq. (\ref{alg}) in two important
respects. First, the fact that the Robertson-Walker universe has fewer
symmetries than de Sitter space gives additional terms in the
semiclassical Einstein equations. Secondly, we will include a
classical stress tensor source representing mixed matter and
radiation. We also set $\L_o =0$.                      
The trace of the semiclassical Einstein equations,
expressed in terms of dimensionless variables, then takes the form of
a simple generalization of Eq. (\ref{dimalg}):
\bea
\label{dimalg3}
r(1+\xib y) \ln \mid 1+\xib y \mid &-& r\frac{\xib y}{1+\xib y} \left\{
1 +{3\over 2}\xib y +\ha y^2
\left(\xib^2 - {1\over 1080}\right)+v \right\} \nn \\
& &= 2\pi\left(y+ 
\frac{T}{2m^2\k_o}\right),
\tea
where $T$ is the trace of the classical stress tensor, and $v$ is a
quantity which vanishes in de Sitter space:
\be
v = \frac{1}{180 m^4}\left({1\over 4}R^2 - R_{\m \n}R^{\m \n}\right).
\te
All quantum contributions are grouped in the LHS of
Eq. (\ref{dimalg3}), while its RHS contains classical terms.

Consider now the full semiclassical Einstein equations with a
classical stress tensor source representing mixed matter and
radiation,
\be
\label{einst}
\einu= \frac{1}{2\k_o}\left(\left(\rho_m+ {4\over 3}\rho_r\right)
u^{\m}u^{\n} + {1\over 3} \rho_r
\metu\right) + {\cal O}(\hbar),
\te 
where $\rho_m$ and $\rho_r$ represent the matter and radiation energy
densities, respectively, and ${\cal O}(\hbar)$ represents the quantum
contributions. The equation above implies
\bea
\label{defn}
y &\equiv& \frac{R}{m^2} =  \frac{\rho_m}{2\k_o m^2} +
\frac{R_Q}{m^2}, \\
\label{defn1}
v &=& - \frac{(\rho_m + (4/3) \rho_r)^2}{960 m^4 \k_o^2} + {\cal
O}(\hbar),
\tea
where $R_Q$, as defined by Eq. (\ref{defn}), is of order $\hbar$. 
In a treatment perturbative in $\hbar$, we replace $y$ and $v$ in the 
LHS of Eq. (\ref{dimalg3}) by their
classical values using the above equations, because the factor $r$ is
of order $\hbar$. This treatment is valid as long as the LHS of
Eq. (\ref{dimalg3}) is small with respect to the term $\pi T/ (m^2
\k_o)$ in the RHS. In the RHS of Eq. (\ref{dimalg3}) we keep the quantum-vacuum
contribution to $y$ as well (i.e. the term $R_Q/m^2$). Furthermore, 
in a Robertson-Walker cosmology with metric
\be
\label{rwmet}
ds^2 = -dt^2 + a(t)^2 \left(\frac{dr^2}{1-kr^2} +r^2 d\O ^2\right),
\te
where $k=-1, +1$ and $0$ denote spatially open, closed and flat
universes respectively,
we have $\rho_m \propto a^{-3}$ and $\rho_r \propto a^{-4}$. It is
then 
more convenient to express $y$ and
$v$ in terms of the present matter and radiation densities
$\rho_{m_0}$ 
and $\rho_{r_0}$, and the redshift
$z$. We therefore introduce new dimensionless variables $d_m$ and
$d_r$,  given by
\bea
\label{defn2}
e^{-d_m} &\equiv & \frac{\rho_{m_0}}{m_{Pl}^2 \k_o}, \\   
\label{defn3}
e^{-d_r} &\equiv & \frac{\rho_{r_0}}{m_{Pl}^2 \k_o},  
\tea
and the redshift $z$ given by
\be
\label{red}
1+z = \frac{a_0}{a},
\te
where $a_0$ is the scale factor at the present time. Furthermore, owing to the fact that we will study the
effect of very light masses, we also redefine the ratio $m^2 / m_{Pl}^2$: 
\be
\label{rat}
r = e^{-S}.
\te
Eq. (\ref{dimalg3}) can now be rewritten by substituting for $y$ and
$v$ using Eqs. (\ref{defn}) and (\ref{defn1}) with
 the redefined variables of Eqs. 
(\ref{defn2}), (\ref{defn2}), (\ref{red}) and (\ref{rat}) above. 
The resulting equation, correct to leading order in
$\hbar$, gives an expression for the quantum correction to the scalar curvature, $R_Q$, as a function of
redshift. Dividing by $R_{cl} = \rho_m/(2\k_o)$, we obtain
\bea
\label{imprat}
\frac{R_Q}{R_{cl}} &\equiv & \frac{R_Q}{\rho_m /(2\k_o)} \nn \\
&=& \frac{e^{d_m-2S}}{\pi (1+z)^3}\left\{ \left(1 +\ha \xib (1+z)^3 e^{S-d_m}\right) \ln \mid 1 +\ha \xib
(1+z)^3 e^{S-d_m}\mid \right.\nn \\
& & - \ha \frac{\xib (1+z)^3 e^{S-d_m}}{1 + \ha \xib (1+z)^3 e^{S-d_m}}\left\{ 1+ {3\over 4} \xib
(1+z)^3 e^{S-d_m} +{1\over 8} \left(\xib^2 - {1\over 108}\right) (1+z)^6 e^{2(S-d_m)}\right.\nn \\
& &~~\left. \left.-{1\over
480}(1+z)^7
e^{2S-d_r-d_m}- {1\over 960}(1+z)^8
e^{2(S-d_r)}\right\}\right\}.
\tea

We wish to find the redshift ranges for which the quantum contribution to the scalar curvature is
significant in comparison with the classical contribution
(i.e. $R_Q/R_{cl}$ of order $1$). The values of $d_r$ and $d_m$ in
Eq. (\ref{imprat}) are determined by 
 the radiation and matter energy densities at the present time. Black body radiation at a temperature of
$2.726 \, K$ gives a radiation energy density $\rho_{r0} \simeq 7.81 \times 10^{-34} {\rm g/cm^3}$ \cite{mather,kolb}.
This gives a
value for  $d_r \simeq 288.06$. The matter density is not known to
good precision. With the conservative estimates
$\O_0 > 0.1$ for the ratio of the present matter
density to the critical density,
and 
$H_0 > 50$km/(s\,Mpc) for the Hubble constant at the present time, we
obtain  $\rho_{m0}>4.70 \times 10^{-31} {\rm
g/cm^3}$ for the matter density. This gives $d_m < 281.66$
for the exponent of Eq. (\ref{defn2}). 

For very light masses ($S>150$), numerical investigations of the ratio $R_Q/R_{cl}$ as a function of
redshift $z$, using Eq. (\ref{imprat})
with $\mid \xib \mid \simeq 1$, reveal that there are two distinct regimes for which this ratio is close to or larger than 1.
 The first such regime occurs at extremely high redshifts ($z > 10^{26}$), close to the GUT scale. 
[In the standard cosmology, the Planck era occurs at a redshift of about $10^{31}$, while the GUT era sets 
in at a redshift of about $10^{26}$.] It is expected that  quantum effects would play a significant
role at such high redshifts.
The {\it second}, unexpected regime for which the quantum-vacuum terms become large occurs at relatively low
redshifts. This
second regime exists only for $\xib<0$, and occurs when the factor $1+(1/2)\xib (1+z)^3 e^{S-d_m}$
in Eq. (\ref{imprat}) approaches zero. This corresponds to values of
$z$ near a redshift
$z_j$ given by 
\be
\label{signi}
(1+z_j)^3 =
-2\xib^{-1}e^{d_m-S}. 
\te
In between these early and late regimes, the quantum contribution to the
scalar curvature is
extremely small and slowly varying, and the evolution of the universe is 
well-approximated by its classical
evolution. 

Numerical investigation of 
the  behavior of the ratio $R_Q/R_{cl}$ for values $\xib=-1$ and $d_m
\simeq 280$, and for very light masses ($S>150$), further reveals that
in the late regime of significance of quantum vacuum terms 
(i.e. near $z=z_j$), 
the
scalar curvature sharply increases as $z \rightarrow z_j$. 
Since $z_j$ depends on $S$ (from Eq. (\ref{signi})), and therefore on
$m$, the value of $m$ dictates the value of $z_j$ at which quantum
vacuum terms can become significant. As we shall see later, an
ultralight mass can lead to quantum vacuum effects becoming
significant at roughly half the age of the universe.
However, it is important to note that when the scalar curvature begins
its rapid increase near redshift $z_j$, 
 quantum effects begin to dominate and  
the perturbative analysis itself breaks down. 

For our purposes, we will only need the result that 
 the ratio $R_Q/R_{cl}$ tends to rapidly increase as $z \rightarrow
z_j^{+}$. The total scalar curvature $R$, which can be written as
\be
R = R_{cl}\left(1+\frac{R_Q}{R_{cl}}\right),
\te
is thus a product of a decreasing function ($R_{cl}$) and an
increasing function. For $z \gg z_j$, $R \simeq R_{cl}$ decreases, and
as $z \rightarrow z_j^{+}$, the quantity $R_Q/R_{cl}$ tends to
suppress the further decrease of $R$.

It will be useful to  have an estimate of $z_j$ relative to the
redshift at matter-radiation equality, $z_{eq}$. The quantity
$z_{eq}$ is defined by the condition of equality of matter
and radiation energy densities:
\be
\label{equal}
\rho_{m0}(1+z_{eq})^3 = \rho_{r0}(1+z_{eq})^4.
\te
Combining Eqs. (\ref{signi}), (\ref{equal}), (\ref{defn2}) and
(\ref{defn3}), we get 
\be
\label{jeq}
\frac{1+z_j}{1+z_{eq}} = \left(\frac{2}{-\xib}\right)^{1/3}e^{4d_m/3 -
d_r-S/3}.
\te 
As earlier, we suppose that $d_r \simeq 288.06$ and $d_m <
281.66$. Also, we will additionally assume that $S > 278$.\footnote{This
 assumption will be justified in the next section, where we find
the value of the mass (and therefore $S$).} With these ranges, we obtain
\be
\frac{1+z_j}{1+z_{eq}} < \left(\frac{2}{-\xib}\right)^{1/3}\, 5.61
\times 10^{-3}.
\te
Thus, if $-\xib$ is of order $1$, we have $z_j \ll z_{eq}$.

It follows, for this range of masses, that the quantum
corrections become significant  at some time during the matter-dominated stage of
the expansion. 
In Section VII, we  derive a formula for the
cosmic time $t_j$ corresponding to the redshift $z_j$, for a spatially open
matter-dominated universe.

To find the scalar curvature at $t_j$, we use Eqs. (\ref{defn2}) and
(\ref{defn3}) to write
\bea
\label{signip}
(1+z_j)^3 e^{S-d_m} &\equiv& (1+z_j)^3 \frac{\rho_{m0}}{m^2 \k_0}
\nn \\
&=& \frac{\rho_{mj}}{m^2\k_0}\nn \\ 
&=& \frac{2R_{cl}(t_j)}{m^2},
\tea
where $\rho_{mj}$ is the matter density at time $t_j$, and the last
equality in the above equation follows from the classical Einstein
equations. Comparing the above equation with Eq. (\ref{signi}) shows
that 
\be
R_{cl}(t_j) = m^2/(-\xib)= \overline{m}^2.
\te

We will now carry out a second analysis, valid outside the
perturbative regime, which shows that the scalar curvature indeed
approaches a constant value as $z \rightarrow z_j$ (or $t \rightarrow
t_j$).
 
%
%

\subsection{Behavior of the scalar curvature for $t>t_j$}

As $t$ approaches $t_j$, the arguments of the previous subsection show
that the classical scalar curvature of the matter-dominated universe
approaches the value $R_{j} =
 \overline{m}^2$. We will now argue that the scalar
curvature does not decrease further for $t>t_j$, i.e.  
it saturates to a value very
close to, but slightly larger than
$R_j$. That is, we will show that there exist approximately de
Sitter-like solutions to Eq. (\ref{dimalg3}) for which the scalar
curvature does not change significantly, although the matter density
keeps decreasing, eventually approaching zero at very late times. 
We find that, for matter densities less than $\rho_{mj}$  there 
exist de Sitter type solutions  of Eq. (\ref{dimalg3}) of the
form
\be
\label{nonper}
R = \overline{m}^2(1-\epsilon)
\te
such that $\mid \e \mid \ll 1$, and $R$ is an extremely slowly varying function
of the matter density. We will assume throughout that the mass $m$ is
very light ($r \ll 1$), and that $-\xib$ is positive and of order $1$. 

To show that Eq. (\ref{nonper}) is indeed a solution, we 
substitute Eq. (\ref{nonper}) in Eq. (\ref{dimalg3}) and assume $v \ll
1$, so that one can effectively ignore $v$. Then Eq. (\ref{dimalg3})
takes the form
\bea
\label{moddim}
\e \ln \mid \e \mid &-& \frac{\e -1}{\e}\left\{1+ {3\over 2} + {1\over
2}(1-\e)^2 (1-(1080\xib^2)^{-1})+ v\right\} =\nn \\
& &~~\frac{2\pi}{r}\left(\frac{\e-1}{\xib} + \frac{T}{2m^2\k_o}\right).
\tea
Define
\be
\label{defnd}
\d \equiv \frac{T}{2m^2\k_o} - \xib^{-1} =
-\frac{\rho_m}{2m^2\k_o}-\xib^{-1}.
\te  
If $\rho_m = \rho_{mj}$, where $\rho_{mj}$ is found from
Eqs. (\ref{signi}) and (\ref{signip}), $\d =0$; and if $\rho_m =0$, $\d =
-\xib^{-1}$. We find that $\mid \e \mid$ is always small within
this range of $\d$ values. In fact, we  show that it is small for
any positive value of $\d$. In the  small $\mid \e \mid$ approximation,
Eq. (\ref{moddim}) becomes
\be
\label{quadra}
\e^{-1}\left(-\frac{1}{2(1080)\xib^2} +v\right) =
\frac{2\pi}{r}\left(\frac{\e}{\xib} + \d\right).
\te
To estimate $v$, we assume that it takes a value in between its value
for a classical matter-dominated universe at time $t_j$, and 
a de Sitter universe, for which
$v=0$. In a classical matter-dominated universe, $v$ is given by
Eq.(\ref{defn1}) as
\be
v = - \frac{\rho_m^2}{960 m^4 \k_o^2}.
\te
At time $t_j$, $\rho_m = \rho_{mj} = 2\k_o \overline{m}^2$. Therefore,
at $t_j$, we obtain
\be
v_j = -(240 \xib^2)^{-1}.
\te
We therefore assume that, for $t>t_j$, $v$ lies in the range
\be
\label{vrange}
-(240 \xib^2)^{-1} < v < 0.
\te

It is convenient to define an additional quantity
\be
\beta = \frac{r}{2\pi}\left(v - \frac{1}{2(1080)\xib^2}\right).
\te
For $r \ll 1$, and $v$ given by Eq. (\ref{vrange}), it follows that
$\mid \beta \mid
\ll 1$. One may now solve Eq. (\ref{quadra}) for $\e$. This yields
\be
\e = -\ha \xib \left( \d \pm \left(\d^2 + 4\beta \xib^{-1}\right)\right).
\te
In order to choose the correct sign in the above equation, we will 
require that the scalar curvature approach its classical value
$\rho_m/(2\k_o)$ for large values of the matter density, i.e. for
$\rho_m/(2m^2\k_o) \gg 1$. For large matter densities, $\d$ is a large
negative number. To get the correct value of the scalar curvature, one
must then choose the solution with the minus sign in the above
equation. Even though this value of $\e$ is not small, continuity 
 demands that we keep the solution with the
same sign for all $\d$. Thus we have
\be
\label{so}
\e = -\ha \xib \left( \d - \left(\d^2 - 4 \beta \xib^{-1}\right)\right).
\te
It is clear, for $\d >0$, that $\mid \e \mid$ is always a small
number, reaching a maximum value of $\sqrt{\beta \xib}$ at $\d
=0$ (we have assumed that $r \ll 1$). 

The approximation $\mid \e \mid \ll 1$, which was made in order to
arrive at the solution $(\ref{so})$, is thus valid for $t \geq t_j$.
Within this range of time, the value of $\epsilon$ evolves from 
\be
\e (t_j) = \sqrt{\beta \xib}
\te
when $\rho_m = \rho_{mj}$ ($\d =0$), to
\be
\e (\infty) \simeq -\beta \xib
\te
when $\rho_m =0$ ($\d = -\xib^{-1}$). 
The fractional change in the scalar curvature during
this range of time is given by Eq. (\ref{nonper}) as
\be
\frac{\D R}{R} = \frac{-\overline{m}^2 \D\e}{\overline{m}^2(1-\e)}
\simeq -\D \e, 
\te 
because $\mid \e \mid \ll 1$ during the entire time range. Thus, with
$\D\e = \e (\infty) - \e (t_j)$, we obtain
\be
\mid \frac{\D R}{R} \mid \simeq \sqrt{\beta \xib}\left(1 + \sqrt{\beta
\xib}\right) \ll 1,
\te
because $r \ll 1$.

To summarize, we have argued that, even in the presence of matter,
there exist de Sitter type solutions in which the scalar curvature is
very close to the value $\overline{m}^2$, and is very slowly varying,
as long as the matter density is such that $\d \geq 0$. For large
negative values of $\d$, i.e. large $\rho_m$ in Eq. (\ref{defnd}), 
it is clear that the scalar curvature continuously changes to
the scalar curvature of a matter-dominated universe.

Coupled with the findings of the previous subsection, in which we
showed that quantum effects near $t_j$ tend to prevent the scalar curvature from
decreasing further, these results strongly point to a cosmology in
which a matter-dominated universe transits to a de Sitter-type
universe. In the next subsection, we outline such a model.

\section{Matter dominated expansion leading into accelerated expansion}

The damping of scalar curvature, which begins at a time close to
$t_j$, 
supports the idea that the 
universe undergoes a transition from a    matter-dominated phase to 
a mildly inflationary de Sitter phase.
The arguments of the previous subsection show that the de Sitter phase
is, to good approximation, described by a solution with $\xib<0$ and
$R \simeq \overline{m}^2$, of
the type found in the Section V, and that the presence of
matter does not significantly change such a solution.
Within a rigorous
framework, the matter-dominated and de Sitter phases  must be
joined in a 
sufficiently smooth manner to gaurantee regularity of all
curvature components at the joining point. However, the perturbative analysis of the previous subsection
shows that the quantum effects become significant over a  short
timescale, after which the scalar curvature approaches a
constant value. This effect allows us to
consider, for comparison with observation,  an approximation in which 
an exact classical matter-dominated solution is  joined (at
time $t_j$) to a
de Sitter solution generated by quantum effects. Assuming a
spatially open cosmology ($k=-1$ in Eq. (\ref{rwmet})),   such a model
is represented by the following
scale factor:
\bea
\label{modelo}
a(t) &=& c_0 \sinh^2(\psi/2), ~~~t = \ha c_0 (\sinh \psi - \psi), ~~~
t<t_j \nn \\
a(t) &=& \a^{-1}\sinh \left(\a (t+c_1)\right),~~~t>t_j.
\tea
Here, $\psi$ parametrizes $a(t)$ and $t$ during the matter-dominated
stage. For $t>t_j$, including the present time $t_0$, 
 the universe is in a  
de Sitter phase. 

Of the five parameters, $c_0$, $\a$, $t_j$, $c_1$ and the
present cosmic time $t_0$, that characterize the model based on
Eq. (\ref{modelo}), not all are independent.
The scalar curvature at the time of joining, $R_j$, must be equal to the
constant scalar curvature during 
the later, inflationary phase, and is determined by the {\it single}
scale 
$\overline{m}
\equiv m/\sqrt{-\xib}$. This requirement and  the requirement
of continuity of the scale factor at the joining point, constitute two
constraints on the
five parameters. The remaining three parameters can, for convenience,
be taken to be (i) the present Hubble constant $H_0$, (ii) the present
ratio of matter density to critical density $\O_0$, and (iii) the mass
scale $\overline{m}$.
Here, we express the five parameters defining the model of
Eq. (\ref{modelo}) in terms of these three basic parameters.

First, the scalar curvature during the later, de Sitter phase is given
by $12 \a^2$. Setting this equal to $\overline{m}^2$, as required by
the de Sitter solutions of Section V, we get the relation
\be
\label{alp}
\a = \frac{\overline{m}}{\sqrt{12}}.
\te

The scalar curvature during the matter-dominated phase is given by $R
= 3 c_0 a^{-3}$. The classical Einstein equations, which hold during
the matter-dominated phase, thus imply
\be
\label{co}
3c_0  = (8\pi G) \rho_{mj} a_j^3 = (8\pi G) \rho_{m0} a_0^3,
\te
where $a_j$ and $a_0$ represent the scale factor at $t_j$ and the
present time $t_0$, respectively, and $\rho_{mj}$ and $\rho_{m0}$ are
the corresponding matter densities. The second equality in the above
equation follows from the fact that $\rho_m a^3$ is  constant during
the evolution, a consequence of conservation of the matter
stress-energy.

Thus, Eq. (\ref{co}) gives
\be
\label{ratio}
c_0/a_0^3 = \O_0 H_0^2,
\te  
where $\O_0 = 8\pi G\rho_{m0}/(3H_0^2)$ is the present ratio of matter
density to critical density.

The Hubble constant at the present time is given by Eq. (\ref{modelo})
as
\bea
H_0 &=& \a \coth (\a (t_0 + c_1))\nn \\
&=& a_0^{-1}\sqrt{1+a_0^2 \a^2},
\tea
where $a_0$ is the scale factor at the present time $t_0$. We solve
for $a_0$, and use Eq. (\ref{alp}) to get
\be
\label{az}
a_0 = (H_0^2 -\overline{m}^2/12)^{-1/2}.
\te
Combining Eqs. (\ref{ratio}) and (\ref{az}), we obtain 
\be
\label{cot}
c_0 = \O_0 H_0^2 (H_0^2 -\overline{m}^2/12)^{-3/2}.
\te

We may obtain the scale factor at time $t_j$, $a_j$, by requiring that
the scalar curvature of the matter-dominated phase approach the value
$\overline{m}^2$ as $t\rightarrow t_j^{-}$. This condition yields
\be
\label{aj}
a_j^3 = \frac{3 c_0}{\overline{m}^{2}}. 
\te
Substituting for $c_0$ from Eq. (\ref{cot}), we then have
\be
\label{ajt}
a_j = (3\O_0 H_0^2/\overline{m}^2)^{1/3}(H_0^2
-\overline{m}^2/12)^{-1/2}.
\te

To obtain $t_j$, we use the matter-dominated solution in
Eq. (\ref{modelo}) to get
\be
\psi_j = 2 \sinh^{-1}\sqrt{a_j c_0^{-1}}
\te
and 
\bea
\label{teej}
t_j &=& \ha c_0 (\sinh \psi_j - \psi_j) \nn \\
&=& c_0 \left( \sqrt{(a_j c_0^{-1})(1+a_j c_0^{-1})} -
\sinh^{-1}\sqrt{a_j c_0^{-1}}\right).
\tea
To obtain $c_1$, we use the de Sitter solution in Eq. (\ref{modelo})
to get
\be
\label{con}
c_1 = \alpha^{-1} \sinh^{-1}(\a a_j) - t_j.
\te

Finally, to obtain the present cosmic time $t_0$, we again use the de
Sitter solution, which yields
\be
\label{to}
t_0 = \a^{-1}\left(\sinh^{-1}(\a a_0) - \sinh^{-1}(\a a_j)\right) +
t_j.
\te
All parameters in the model are now expressed in terms of
$\overline{m}$, $H_0$ and $\O_0$. 
 
In the next section, we will compare the predictions of this model to
recent data from high-redshift Type 1a
supernovae \cite{perl}, using magnitude-redshift curves obtained from 
our model. 
 
\section{Comparison of theory with high-$z$ supernovae data;
prediction of particle with mass $\sim 10^{-33}$ eV}
Recent observations of Type 1a supernovae at high redshifts indicate a negative value of the deceleration
parameter at the present time, i.e. an accelerating universe
\cite{perl}. 
Previous attempts to account for this phenomenon invoke a cosmological
constant, or a classical scalar field, quintessence \cite{caldwell}, with 
unusual potentials.  
To explain the observed acceleration effect by means of a cosmological
constant, it must contribute a term to the Einstein equations that is
of the same order of magnitude as that attributed by the present
matter density. On the other hand, 
quintessence models that account for the acceleration effect typically involve
potentials that would give rise to 
nonrenormalizable  quantum field theories.

In the model we present here, the value of $H_0$ is fixed by
low-redshift measurements, while the remaining free parameters
$\overline{m}$ and $\O_0$ are determined by the SNe-Ia data. Once
the mass scale $\overline{m}$ is determined, no fundamental parameters in the
effective action need be chosen to fit the supernovae data. 
Furthermore, the theory 
we work with arises out of a renormalized effective action.

Comparison of our model to SNe-Ia data is achieved by  fitting
calculated magnitude-redshift curves to
the data. The difference between the apparent magnitude ($m$) and absolute magnitude ($M$) of a source is
given in terms of the luminosity distance $d_L$ to the source, by
\be
\label{diffmag}
m-M = 5 \log_{10}\frac{d_L}{{\rm Mpc}} + 25.
\te
The luminosity distance itself is given by \cite{wein}
\be
\label{lum1}
d_L = (1+z)a_0 r_1,
\te
where $a_0$ is the present scale factor, and $r_1$ is the comoving
coordinate distance from a source at redshift $z$ to a detector at
redshift $0$. For Robertson-Walker universes, $r_1$ is given by the
equation
\be
\label{lum2}
\int_0^{r_1}\frac{dr}{(1-kr^2)^{\ha}} = a_0^{-1}\int_0^z
\frac{dz'}{H(z')},
\te
where $k=0, +1, -1$ correspond to flat, closed and open universes,
respectively. For a spatially open universe, the above equation
reduces to
\be
\label{lum3}
\sinh^{-1}r_1 = a_0^{-1}\int_0^z \frac{dz'}{H(z')}.
\te

Consider a universe represented by the model of Eq. (\ref{modelo}),
which is a spatially open matter-dominated universe prior to time
$t_j$, and transits into a de Sitter
universe at $t_j$. Let $z_j$ be the redshift at time $t_j$.
For $z <z_j$, the universe is in a de Sitter phase, with Hubble
constant given by
\bea
\label{hubble}
H &=& \a \coth \a (t+c_1) \nn \\
&=& \sqrt{\a ^2 + \frac{(1+z)^2}{a_0^2}}.
\tea 
Substituting the above into Eq. (\ref{lum3}), and performing the
integration, we obtain, for $z <z_j$,
\be
\label{rol}
r_{1<}(z) = \frac{1+z}{(a_0 \a)^2}\left(\sqrt{1 + (a_0 \a)^2} -
\sqrt{1+\left(\frac{a_0\a}{1+z}\right)^2}\right),
\te
where $r_{1<}(z)$ denotes $r_1(z)$ for $z <z_j$.  For $z >z_j$, the
RHS of 
Eq. (\ref{lum3}) separates into two
contributions:
\be
\label{lum4}
\sinh^{-1}(r_{1>}) = a_0^{-1}\left(\int_0^{z_j} \frac{dz'}{H(z')}+
\int_{z_j}^z  \frac{dz'}{H(z')}\right),
\te 
where $r_{1>}(z)$ denotes $r_1(z)$ for $z>z_j$.

During the matter-dominated phase, the Hubble constant is calculated
as
\bea
\label{hubble2}
H &=& \sqrt{c_0 a^{-3} + a^{-2}} \nn \\
&=& \frac{1+z}{a_0}\sqrt{1+ c_0 a_0^{-1}(1+z)}.
\tea
Using Eq. (\ref{hubble}) for $z'<z_j$ and Eq. (\ref{hubble2}) for
$z'>z_j$, the integrations in Eq. (\ref{lum4}) may be performed to yield 
\be
\label{rog}
r_{1>}(z) = \sinh \left(\sinh^{-1}(r_{1<}(z_j)) + \ln
\left(\frac{(g(z)-1) (g(z_j)+1)}{(g(z)+1) (g(z_j)-1)}\right)\right),
\te
where 
\be
g(z) = \sqrt{1+c_0 a_0^{-1}(1+z)}.
\te
Eq. (\ref{lum1}) gives the luminosity distance, $d_{L1}$, for this model, as
\bea
\label{lumd1}
d_{L1} &=& (1+z)a_0 r_{1<}(z), ~~~z<z_j \nn \\
&=& (1+z)a_0 r_{1>}(z), ~~~z>z_j. 
\tea
 
%
For comparison, the luminosity-distance-redshift relation
for a matter-dominated Robertson-Walker  universe with zero 
cosmological constant and ratio $\O_0$ of present matter
density to critical density, is
\be
d_{L2}(\O_0, z) = 2 H_0^{-1}\O_0^{-2}\left(\O_0 z +(\O_0-2)(\sqrt{1+\O_0 z} -1)\right),
\te
and for a spatially flat matter-dominated ($\O_0 =1$) universe,
\be
d_{L2}(1, z) = 2 H_0^{-1}\sqrt{1+z}(\sqrt{1+z}-1).
\te

\begin{figure}[hbt]
\centering
\leavevmode
\epsfysize=5.in\epsffile{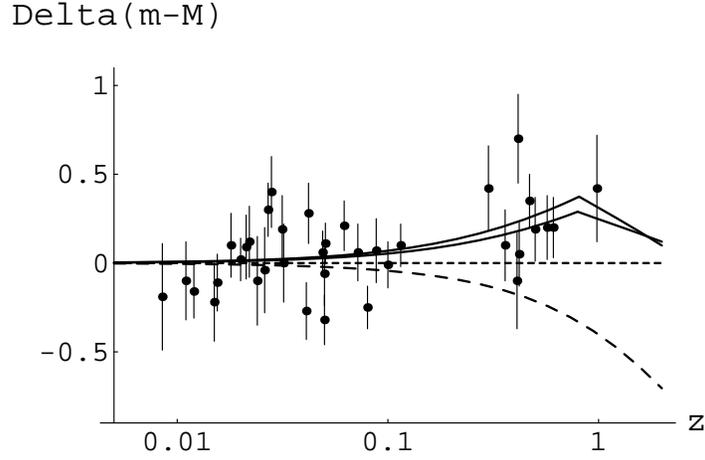}
\caption{A plot of the difference between apparent and absolute
magnitudes, 
as functions of redshift $z$, normalized to an open universe with
$\O_0 =0.2$ and zero cosmological constant. The points with vertical
error bars represent SNe-Ia data obtained from
Ref.[5]. 
The two
solid curves represent the values a) $\overline{m} = 3.7 \times
10^{-33}$ eV and $\O_0 = 0.4$ (upper solid curve), and b) $\overline{m}
= 3.2 \times 10^{-33}$ eV and $\O_0 = 0.3$ (lower solid curve). 
The horizontal dashed line
represents an open universe with $\O_0 = 0.2$, and the dashed line
curving downward represents a matter-dominated flat universe. Smaller
values of $\O_0$ also would fit the data 
(see text after Eq. (112)).}
\label{figu12}
\end{figure}

It is convenient (see Eq. (\ref{diffmag})) to define 
\be
\label{diffmnorm}
\D (m-M)(z) = 5 \log_{10}\left(\frac{d_{L}(z)}{d_{L2}(0.2, z)}\right) .
\te 
Fig. \ref{figu12} is a plot of $\D(m-M)$ vs. $z$, along with a plot of
SNe-Ia data acquired  from Ref. \cite{perl}. The two solid curves
represent $d_{L}(z) = d_{L1}(z)$.
In plotting this quantity, all parameters appearing in $d_{L1}$ have
been expressed in terms of the three basic parameters $\overline{m}$,
$H_0$ and $\O_0$, using the relations derived in the previous
section. 
Also, the
value of $H_0$ has been set to $65 {\rm km}/({\rm s\,Mpc})$. Thus there are two
quantities, $\overline{m}$ and $\O_0$, which parametrize the solid
curves. The two curves shown in the figure give a reasonable fit to
the data, and correspond to a)
$\overline{m}= 3.7 \times 10^{-33}$ eV and $\O_0 =0.4$ (higher solid
curve), and b) $\overline{m} = 3.2 \times 10^{-33}$ eV and $\O_0 =0.3$
(lower solid curve). 

The general features of a family of curves parametrized by
$(\overline{m}, \O_0)$ are as  follows. For a fixed value of
$\overline{m}$, decreasing $\O_0$ has the sole effect of increasing the
redshift at which the transition occurs, i.e. a smaller value of $\O_0$
will move the transition further from the present time. Thus we cannot
rule out the possibility that, with more observations at higher
redshift, a better fit to the data could be obtained with lower values
of $\O_0$. However, the data do not allow the joining points in
Fig. (\ref{figu12}) to occur at smaller $z_j$, so the values of 
$\O_0$ shown in the plot do 
 represent a rough upper bound on $\O_0$ in our model, and lead to
the conclusion stated earlier, $\O_0 < 0.4$.
As is well known, an {\it early} inflationary epoch would explain why
$\O_0$ is not very far from $1$. 

For a fixed value of $\O_0$, increasing $\overline{m}$ has the effect
of shifting the curves up, as well as increasing somewhat the redshift
at which the transition occurs.

The two dashed curves in Fig. (\ref{figu12}) are shown for comparison,
and represent c) $d_{L} =
d_{L2}(0.2, z)$ (horizontal dashed line), i.e. an open
matter-dominated universe with $\O_m =0.2$, and d) $d_{L} = d_{L2}(1,
z)$ (lower dashed curve), i.e. a spatially flat matter-dominated universe.

\section{The age of the universe}

As stated earlier, the only fundamental scale that enters into the
effective action of the model presented here is $\overline{m}$. 
Nevertheless, as we show now, the fit of our model to supernovae data
predicts reasonable values for the age of the universe $t_0$.

The relations derived in Section VII, leading up to 
Eqs. (\ref{teej}) and (\ref{to}), give $t_j$ and $t_0$ in terms of
$\overline{m}$, $\O_0$ and $H_0$. For $H_0 = 65\, {\rm km}/({\rm s\,Mpc})$,
$\overline{m}= 3.7 \times 10^{-33}$ eV, and $\O_0 =0.4$ (upper solid
curve in Fig. \ref{figu12}), we obtain
\bea
t_j &=& 5.66 \times 10^9 {\rm years} \\
\label{timeso}
t_0 &=& 1.34 \times 10^{10} {\rm years}.
\tea

For $H_0 = 65\, {\rm km}/({\rm s\,Mpc})$, 
$\overline{m} = 3.2 \times 10^{-33}$ eV and
$\O_0 = 0.3$, we obtain
\bea
t_j &=& 6.03 \times 10^9 {\rm years} \\
\label{timest}
t_0 &=& 1.33 \times 10^{10} {\rm years}.
\tea

Therefore, in both cases a reasonable value of roughly $13$ billion
years is obtained for the age of the universe. More data at
higher redshifts may lower the value of $\O_0$ in our model, which
would further increase the age of the universe.

\begin{figure}[hbt]
\centering
\leavevmode
\epsfysize=6.in\epsffile{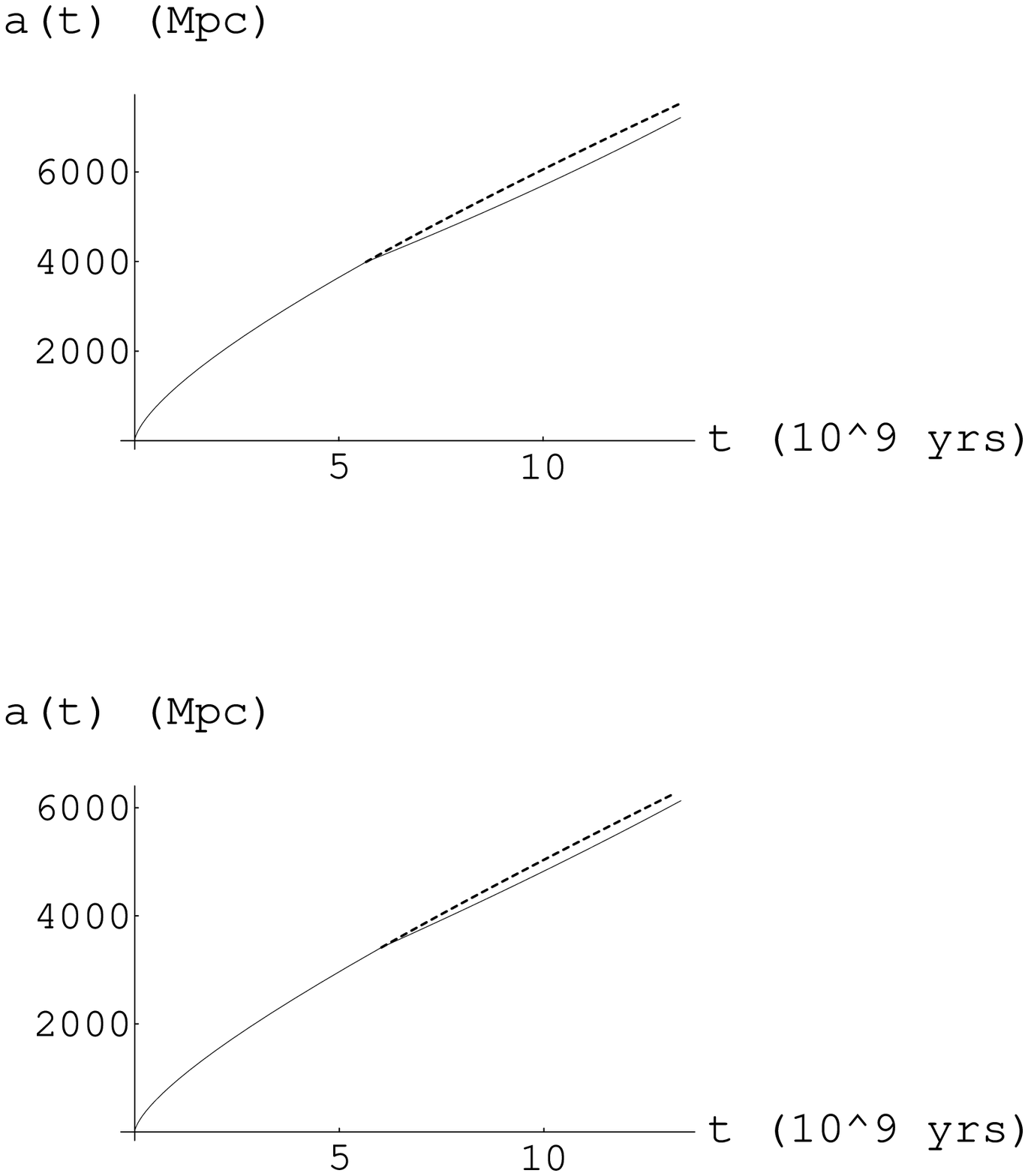}
\caption{Two plots of the scale factor versus time 
 for a spatially open model universe in
which an initially spatially open matter-dominated cosmology evolves to a
de Sitter solution. The parameters for the top model are $\overline{m} =
3.7 \times 10^{-33}$ eV and $\O_0 =0.4$, and for the bottom model,
 $\overline{m} =
3.2 \times 10^{-33}$ eV and $\O_0 =0.3$.
The dashed curves represent a continuation of the open
matter-dominated phase.}
\label{figu13}
\end{figure}

Fig. \ref{figu13}  contains plots of the scale factor
versus cosmic time for the two solid curves of Fig. \ref{figu12}. In
each case, the open matter-dominated universe that transits to the de
Sitter phase is shown continued as a dashed curve, for comparison.

\section{Conclusions}

In summary, we showed that a model in which a transition occurs from
a matter-dominated to a de Sitter expansion, fits the SNe-Ia data. We
call such a model, a transitionary universe. Also, we have proposed a
free quantum field theory effective action, Eq. (\ref{ful}), 
in which such a transition
evidently occurs. The existence of a particle of very
low mass would cause the universe to make a transition from the
matter-dominated to a new de Sitter stage. In our model, one can say
that we are now observing the mass scale of this particle through the
SNe-Ia data.

Models involving interacting fields may also give a
transitionary universe, and such models would be natural extensions of
the free field model presented here as the simplest case. 

We emphasize once again that the solutions to the gravitational field
equations  we 
obtain, in particular
the de Sitter solutions for $\xib<0$, exist without the necessity of a non-zero
cosmological constant term in the effective action. Furthermore, these
solutions
 are fairly insensitive to the
presence of a cosmological constant term. In this manner,
our model does not suffer from the problem of fine-tuning of the
cosmological constant, which exists in mixed matter and vacuum energy
models.
 
The  present matter density 
 and the predicted age of the universe agree  well with the
current estimates. As in other models, a value of $\O_0$ not far from $1$
may result from a period of early inflation.
Further constraints on $\O_0$ and $\overline{m}$ in our model could
result from comparison to cosmic microwave background data, as well as from the
time-temperature relationship during nucleosynthesis. We hope to carry out
such a comparison in the future.

Finally, we would like to mention that the $R$-summed form of
the effective action could have consequences for early universe
cosmology as well. In particular, the existence of an imaginary term
in the effective action, implying particle creation effects, could
play a role in the exit from an inflationary universe. In future work,
we plan to pursue these ideas 
as well as to carry out a dynamical calculation giving the details of the 
transition between the matter-dominated stage and the later de Sitter stage
of the expansion.

\noindent {\bf Acknowledgements}

\noindent This work was supported by NSF grant PHY-9507740.

\newpage

\appendix
\section{Divergences in the Effective Action as $M^2 \rightarrow 0^+$}

Consider  the one-loop effective action in Eq. (\ref{effser}) below
threshold, i.e. $M^2 >0$.
This action is 
then a real series which
is divergent as $M^2 \rightarrow 0$. The part of the first term in
this series, involving $M^4$,  vanishes
in this
limit. However, subsequent terms, involving $\overline{f}_l$ for $l
\geq 2$, are all divergent  as $M^2
\rightarrow 0$. All terms containing $\overline{f}_3$ and higher
orders correspond to an asymptotic  series
in inverse powers of $M^2$
and therefore do not give a valid expansion for small values of
$M^2$. The behavior of  this
expansion as $M^2
\rightarrow 0$ is thus unphysical and does not
seem to be a cause for concern\footnote{It is possible that all higher
order terms sum to give a  finite
contribution as $M^2 \rightarrow 0$. An example of such a situation is
the series  $(1-x^{-1})^{-1} =
\sum_{n=0}^{\infty} x^{-n}$, which is a valid expansion for $x >1$. As
$x \rightarrow 0$, every term  on the 
right hand side is divergent, although the left hand side
vanishes.}. However, the term  $\overline{f}_2 \ln
(M^2)$ is also divergent in this limit and
is physically required for renormalization of ultraviolet (UV)
divergences and to obtain the  trace anomaly.
It is therefore of interest to examine in detail the divergent
behavior of  this term as $M^2 \rightarrow 0$. 
We will show here that this divergence, although infrared in nature,
can be  absorbed into the
ultraviolet divergences of the theory by a procedure similar to the
way  infrared divergences at $m=0$ are
handled in the usual one-loop effective action \cite{dew}. To this
end, we  will consider the 
one-loop effective action (\ref{effser}), truncated up to the terms
involving  $\overline{f}_2$. We will work 
within the
dimensional regularization scheme, which is more 
useful than the zeta function scheme in this context because the
divergent  terms are explicitly displayed.

First, Eq. (\ref{effo}) implies the following proper time representation of 
the effective action
\cite{schw,dew}:
\bea
W^1 &=& -{i\over 2} {\rm Tr} \intf ds s^{-1} {\rm e}^{-isH} \nn\\
&=& {1\over 2} (\mu^2)^{2-D/2}\intx \intf ids (is)^{-1}(4\pi
is)^{-D/2} {\rm e}^{-is(M^2
-i\epsilon)}\overline{F}(x,x, is).
\tea
The truncated one-loop effective action before renormalization is
obtained, in four dimensions, by keeping  the
first three terms in 
a power series expansion of $\overline{F}$ in Eq. (\ref{sdw}). These
terms include  UV-divergent
contributions to the effective
action arising from the behavior of the integrand near $s=0$. All
higher order terms are  UV-finite.

Thus we obtain for the truncated one-loop effective action
\be
\label{tr}
W^1_{\rm trun} = \ha (4\pi)^{-D/2}(\mu^2)^{2-D/2}\intx \intf ids
(is)^{-D/2-1}{\rm e}^{-is(M^2
-i\epsilon)}(1
+(is)^{2}\overline{f}_{2}),
\te 
where we have set $\overline{f}_1 =0$ without loss of generality.
The above equation has a finite piece corresponding to the sum over
all powers of $s$ higher than $2$  in
the
expansion of $e^{-is\xib R}$ in the integrand. Therefore $W^1_{\rm
trun}$ as defined above differs  from
what is usually regarded as the divergent part of the effective action
by this finite piece.  Here, we need
to keep this extra piece in order to properly take the limit $M^2
\rightarrow 0$. 
 
In performing dimensional regularization about $D=4$, 
it is convenient to define $2\d = D-4$. 
One can perform  the proper time integral in Eq. (\ref{tr}), to get
\be
W^1_{\rm trun} = (32\pi^2)^{-1}\intxf \left\{ \mu^4
\left(\frac{M^2}{\mu^2}\right)^{2+\d }\,  \G (-2-\d )  +
\overline{f}_2
\left(\frac{M^2}{\mu^2}\right)^{\d }\,\G
(-\d)\right\}.
\te
Expanding the exponents and the Gamma functions about $\d =0$
($D=4$),
we get
\bea
W^1_{\rm trun} &=& (32\pi^2)^{-1}\intxf \left\{M^4
\left(\frac{1}{4-D}-\frac{\g }{2} +{3\over 4}-  \ha \ln
(M^2/\mu^2)\right)\right. \nn \\
& &~~~+
\left.\overline{f}_2 \left(\frac{2}{4-D} -\g - \ln
(M^2/\mu^2)\right)\right\} +  {\cal O}(D-4).
\tea
The  expression above 
already indicates that the divergence as $M^2 \rightarrow 0$ may be
absorbed into the UV-divergence  as $D
\rightarrow 4$ in the
coefficient of $\overline{f}_2$. 
To see this explicitly, we may set  $M^2 =0$ in
Eq. (\ref{tr}) at the outset, and replace the upper limit of the
$s$-integration by  some
large number $T$ to regularize the integral (infrared
regularization). We can then examine the  divergent
behavior as $T \rightarrow
\infty$. We therefore have 
\be
W^1_{\rm trun} = (32\pi^2)^{-1}(\mu^2)^{2-D/2}\intx \int_0^T ids
(is)^{-D/2-1}{\rm e}^{-s\epsilon}(1
+ (is)^{2}\overline{f}_{2}),
\te 
which may be evaluated in terms of the incomplete Gamma functions $\g(\a, x)$, 
\bea
W^1_{\rm trun} &=& (32\pi^2)^{-1}\intxf 
\left\{ i^{-D/2}(\e/\mu^2)^{D/2} \,\g (-D/2, \e T)\right.  \nn \\
& &~~+\left.i^{-D/2+2}\overline{f}_{2} (\e/\mu^2)^{D/2-2}\, \g (2-D/2,
\e T) \right\}.
\tea
We may now take the limit $\e \rightarrow 0$ of the above expression
by using the power series  expansion
\cite{grad}
\be
\g (\a, x) =\sum_{n=0}^{\infty}\frac{(-1)^n x^{\a +n}}{n! (\a +n)}.
\te
This yields, after some simplification
\be
\label{infra}
W^1_{\rm trun} = (32\pi^2)^{-1} \intxf \left\{ \ha T^{-2} +
\overline{f}_2 \left(\frac{2}{4-D} +  \ln(\mu^2
T)\right) + {\cal O}(D-4) \right\}.
\te
As $T \rightarrow \infty$, the first term in Eq. (\ref{infra}) drops
out, leaving behind a term  proportional
to $\overline{f}_2$ which diverges logarithmically in this
limit. However, this divergence can be absorbed 
into the
UV-divergence as $D\rightarrow 4$.
  
We therefore find that the logarithmic divergence in the effective
action in the ``infrared'' limit  
$T\rightarrow \infty$ may be canceled by  counterterms of the same
geometric form as the ones introduced 
into the bare gravitational
action to cancel the UV divergences as
$D \rightarrow 4$, i.e. a counterterm proportional to $\overline{f}_2$
is required here. The infrared 
problem in the
partially summed form of the effective action is thus 
handled in a way similar to the infrared divergence in the usual
effective action at $m=0$, where a 
counterterm proportional to $f_2$ is required.

The analysis above may be carried out in a similar manner even when we
do not set $M^2 =0$ at the very 
beginning but rather let $M^2$ tend to zero from positive
values. Eq. (\ref{infra}) is recovered at the end 
of the calculation.

\section{Variations of curvature invariants}

Here we will list the variations of the curvature invariants that
occur in the effective action    
in Eq. (\ref{ful}). They are as follows:
\be
\d (\intxf R) = - \intxf \,\d g_{\m \n}\, G^{\m \n }
\te
\be
\d (\intxf\, R^2) = \intxf \,\d \met\, \left\{ \ha \metu R^2 - 2R \ricu 
+ 2R^{;\n \m} - 2\metu \Box R\right\}
\te
\bea
\lefteqn{\d (\intxf \ln \mid M^2/m^2 \mid ) = \intxf\, \d \met \,
\left\{ -\xib M^{-2} R^{\m \n } - \xib  
g^{\m \n}\left(2M^{-6} \xib^2 R_{;\a}R^{;\a}\right.\right.}\nn \\
& & - \left.M^{-4}\xib \Box R\right) 
+ \xib \left(2M^{-6}R^{;\m }R^{;\n } - M^{-4}\xib R^{;\n \m }\right)
 +\left. \ha g^{\m \n }\ln \mid
M^2/m^2 \mid \right\}
\tea
\bea
\lefteqn{\d (\intxf \,R \ln \mid M^2/m^2 \mid) = \intxf \,\d \met\,
\left\{ -\einu \ln \mid M^2/m^2 \mid 
-\xib M^{-2} R R^{\m \n} \right.}\nn \\
& &+ \xib \frac{m^2 +M^2}{M^4}(R^{;\m \n} - \metu \Box R) 
-\left.\xib^2 \frac{2m^2 +M^2}{M^6} (R^{;\m}R^{;\n} - \metu R_{;\a
}R^{;\a }) \right\} 
\tea
\bea
\lefteqn{\d (\intxf R^2 \ln \mid M^2/m^2 \mid ) = \intxf \,\d \met
\,\left\{ \ln \mid M^2/m^2 \mid\left(\ha 
\metu R^2 - 2R \ricu \right.\right.}\nn \\
& &+\left.2 R^{;\n \m} \right)
+ \xib M^{-2}\left(-R^2 \ricu +6 R^{;\m}R^{;\n} +4R R^{;\n \m }\right) 
- \xib^2 M^{-4}R\left(6R^{;\m}R^{;\n}\right. \nn \\
& &\left. +\left. R R^{;\n \m}\right) + 2\xib^3 M^{-6}R^2 R^{;\m}R^{;\n}
\right\} 
\tea
\bea
\lefteqn{\d (\intxf \,\overline{f}_2) = {1\over 180} \intxf \,\d
\met\, \left\{3\Box \ricu - R^{;\m \n} - 
\ha
\metu \Box R \right.}\nn \\
& &+ 2 R^{\m \s \n \t}R_{\s \t } + \ha \metu R_{\s \t}R^{\s \t} 
+\left. 2{R^{\m}}_{\s \t \rho} R^{\n \s \t \rho} -\ha \metu R_{\s \t
\rho \l}R^{\s \t \rho \l} 
-4 {R^{\m}}_{\s}R^{\n \s} \right\}
\tea
\bea
\lefteqn{\d (\intxf \Box R\, \ln \mid M^2/m^2 \mid) = \intxf \, \d
\met \, \xib M^{-2} \left\{ -2 \metu 
\Box \Box R + 2(\Box R)^{;\mu \nu} \right.}\nn \\
& & -\ha \metu R^{;\a}R_{;\a} + R^{;\m }R_{;\n} +\xib
M^{-2}\left(4\metu (\Box R)^{;\a} R_{;\a } +2\metu 
(\Box R)^2 - 4 R^{;(\m }(\Box R)^{;\n )} -2\Box R \,R^{;\m \n} \right. \nn \\
& &+ R^{\m \n}R^{;\a }R_{;\a}
 +\left. 2R^{;\a}(\metu \Box - \nabla^{\m}\nabla^{\n})R_{;\a} + 2\metu
R_{;\a \b}R^{;\a \b} 
-2{R_{;\a}}^{\n}R^{;\a \m} \right)\nn \\
& &+2\xib^2 M^{-4}\left(- R_{;\a}R^{;\a }(\metu\Box - \nabla^{\m}\nabla^{\n}) R
+4R^{;(\m}{R_{\a}}^{\n)}R^{;\a} -4 \metu R^{;\a}R^{;\b}R_{;\a \b}\right)\nn \\
& &+\left. 6\xib^3 M^{-6}\left(\metu R_{;\a}R^{;\a} -
R^{;\m}R^{;\n}\right) \right\} 
\tea 
\bea
\lefteqn{\d (\intxf  R^{\a \b}R_{\a \b}\, \ln \mid M^2/m^2 \mid) =
\intxf \, \d \met \,  
\left\{ \ln \mid M^2/m^2 \mid \left( \ha \metu R^{\a \b}R_{\a \b} - 3
R^{\m \a}R^{\n}_{\a} 
\right.\right.}\nn \\
& &+\left.\Box \ricu + \ha \metu \Box R - R^{;\m \n} + R^{\a \n \b
\m}R_{\a \b} \right)+\xib 
M^{-2}\left(-\ricu R^{\a \b}R_{\a \b} + \ricu \Box R - 2R^{\a
(\m}{R_{;\a}}^{\n)} \right. \nn \\ 
& &+\metu R^{\a \b}R_{;\a \b} + 2R_{;\a}R^{\m \n; \a} +\metu
R_{;\a}R^{;\a} - R^{;\m}R^{;\n} -2 R_{;\a}R^{\a 
(\m;\n)} - 2\metu R^{\a \b}\Box R_{\a \b} \nn \\
& &- 2\metu R_{\a \b; \k}{R^{\a \b}}_{;\k} 
+\left.2 {R_{\a \b}}^{;\n \m}R^{\a \b} +2{R_{\a \b}}^{;\m}R^{\a \b;
\n} \right)+\xib^2 M^{-4} 
\left(-\ricu R_{;\a}R^{;\a} - \metu R^{\a \b}R_{;\a}R_{;\b} \right.\nn \\
& &+\left.2 R^{;(\m}R^{\n)\a}R_{;\a} +4\metu R^{\a \b}R^{;\k}R_{\a
\b;\k}+\metu R_{\a \b}R^{\a \b}\Box R -4 
{R_{\a
\b}}^{;(\m}R^{;\n)}R^{\a \b} - R^{\a \b}R_{\a \b}R^{;\m \n} \right)\nn \\
& &+\left.2\xib^3 M^{-6} R^{\a \b}R_{\a \b}\left(R^{;\m}R^{;\n} -
\metu R_{;\a}R^{;\a}\right) \right\}  
\tea
\bea
\lefteqn{\d (\intxf  R^{\a \b \g \d}R_{\a \b \g \d}\, \ln \mid M^2/m^2
\mid) = \intxf \, \d \met \,  
\left\{ \ln \mid M^2/m^2 \mid \left( \ha \metu R^{\a \b \g \d}R_{\a \b
\g \d} \right. \right. }\nn \\ 
& &-2R^{\m \k \g \d}{R^{\n}}_{\k \g \d} - 4 {R^{\b \n \a \m}}_{;\b \a}
+\xib M^{-2}\left(-\ricu \rimt -2 
\metu R^{\a \b \g \d}{R_{\a \b \g \d;\k}}^{\k} \right. \nn
\\
& &- 2 \metu R^{\a \b \g \d; \k}R_{\a \b \g \d;\k} 
+\left.2{R_{\a \b \g \d}}^{;\n \m}R^{\a \b \g \d} + 2 {R_{\a \b \g
\d}}^{;\n} R^{\a \b \g \d;\m} -4 {R^{\b \n\a 
\m}}_{;\a}R_{;\b} -4 R^{\b \n \a \m}R_{;\b \a}\right)\nn \\
& &+\xib^2 M^{-4} \left( 4\metu \rimu {R_{\a \b \g \d }}^{;\k}R_{;\k}
+\metu \rimt \Box R -4\rimu {R_{\a \b \g \d }}^{;(\n} R^{;\m)}
-\rimt R^{;\m \n}\right. \nn \\
& &+\left.4R^{\b \n \a \m}R_{;\b }R_{;\a} \right) 
+ \left.2\xib^3 M^{-6}\left(-\metu R_{;\a}R^{;\a} + R^{;\m}R^{;\n}
\right) \right\}   
\tea

\newpage

\end{document}